\documentclass[preprint]{ptephy_v1}
\preprintnumber{XXXX-XXXX} 
\usepackage{amssymb}
\usepackage{amsmath}
\usepackage{graphicx}
\usepackage{subfigure}
\usepackage{amsthm}
\usepackage{mathrsfs}
\usepackage{indentfirst}
\usepackage[colorlinks,citecolor=blue,urlcolor=blue]{hyperref}
\allowdisplaybreaks
\usepackage{tikz-network}
\usepackage{pgf}
\usepackage{tikz}
\usetikzlibrary{arrows, decorations.pathmorphing, backgrounds, positioning, fit, petri, automata}
\usetikzlibrary{shadows,arrows,positioning}
\RequirePackage{algorithm,algpseudocode}
\theoremstyle{plain}
\newtheorem{Theorem}{Theorem}

\newtheorem{Proposition}{Proposition}
\newtheorem{Definition}{Definition}
\newtheorem{Lemma}{Lemma}
\newtheorem{Assumption}{Assumption}
\newtheorem{Remark}{Remark}
\newtheorem{Example}{Example}
\begin{document}

\title{Distribution-Free Model for Community Detection}


\author{Huan Qing}
\affil{School of Mathematics, China University of Mining and Technology, Xuzhou, 221116, Jiangsu Province, China\email{qinghuan@cumt.edu.cn;qinghuan07131995@163.com}}

\begin{abstract}%
Community detection for unweighted networks has been widely studied in network analysis, but the case of weighted networks remains a challenge. This paper proposes a general Distribution-Free Model (DFM) for weighted networks in which nodes are partitioned into different communities. DFM can be seen as a generalization of the famous stochastic blockmodels from unweighted networks to weighted networks. DFM does not require prior knowledge of a specific distribution for elements of the adjacency matrix but only the expected value. In particular, signed networks with latent community structures can be modeled by DFM. We build a theoretical guarantee to show that a simple spectral clustering algorithm stably yields consistent community detection under DFM. We also propose a four-step data generation process to generate adjacency matrices with missing edges by combining DFM, noise matrix, and a model for unweighted networks. Using experiments with simulated and real datasets, we show that some benchmark algorithms can successfully recover community membership for weighted networks generated by the proposed data generation process.
\end{abstract}
\subjectindex{A13, A50, H40, J71}
\maketitle
\section{Introduction}\label{sec1}
For decades, network analysis appears in various applications in areas such as social science, biological science, physics, compute science and statistics \cite{watts2002identity,girvan2002community,newman2003the,palla2007quantifying,barabasi2004network,guimera2005functional,lusseau2004identifying,newman2001the,newman2002random,airoldi2013multi,ji2016coauthorship,ji2021co}. Overviews of network science and statistical modeling can be found in, for example, \cite{newman2003the,goldenberg2010a}.

Community detection is a popular tool to extract latent structural information of a network in which nodes are partitioned into different communities. The Stochastic Blockmodel (SBM) \cite{SBM} is a classical and widely used model to model unweighted networks. Based on SBM, various meaningful models are developed, for example, the Degree-Corrected Stochastic Blockmodels (DCSBM)\cite{DCSBM} extends SBM by introducing degree heterogeneity to model real-world networks in which nodes have variation degrees; the Mixed Membership Stochastic Blockmodel (MMSB) \cite{MMSB} models networks in which nodes may belong to multiple communities; \cite{MixedSCORE}'s DCMM and \cite{OCCAM}'s OCCAM extends MMSB by considering degree heterogeneity; the ScBM and DCScBM proposed in \cite{DISIM} extend SBM and DCSBM to model directed networks in which edges have directional property. In recent years, based on these models, substantial works related to applications, algorithms, and theoretical frameworks have been developed, for example, \cite{rohe2011spectral,lei2015consistency, SCORE, SPACL,chen2018convexified,zhou2019analysis,DSCORE,zhao2012consistency,joseph2016impact,choi2011stochastic,abbe2015community,abbe2016exact,hajek2016achieving,gao2017achieving}, and the references therein. An overview of recent developments on SBM can be found in \cite{abbe2017community}. In particular, spectral clustering \cite{von2007tutorial} is arguably one of the most widely applied approaches to recover communities for networks generated from SBM and DCSBM \cite{rohe2011spectral,RSC,lei2015consistency,SCORE,SLIM}. Broadly speaking, spectral clustering infers the community membership by applying a clustering algorithm to a matrix formed by eigenvectors of the adjacency matrix or its variants. Spectral clustering is easy to implement and has been shown theoretically consistent \cite{rohe2011spectral,RSC,lei2015consistency,SCORE,joseph2016impact,MaoSVM,su2019strong,SPACL}.

However, a significant limitation of SBM is that it can not model weighted networks. Edge weights are meaningful in networks, and can improve community detection \cite{newman2004analysis,barrat2004the}. To overcome this limitation, in recent years,
some Weighted Stochastic Blockmodels (WSBM) are developed \cite{aicher2015learning, jog2015information,ahn2018hypergraph, palowitch2018significance,xu2020optimal,ng2021weighted}. However, these weighted models always limit the connectivity matrix (defined in Equation (\ref{definP})) to having nonnegative elements or just a few elements compared to the number of communities, another limitation of these weighted models is some of them require the edge weights to follow certain distributions. Though the two-way blockmodels proposed in \cite{airoldi2013multi} can model weighted networks in which edge weights can be negative, it limits the adjacency matrix's elements to following Normal distribution. In general, the above WSBMs can not model weighted networks whose edge weights follow Normal distribution because of their nonnegative requirement on elements of the connectivity matrix. To overcome the shortcomings of these weighted models, this paper proposes the Distribution-Free Model. 

The key contributions in this article are as follows:

(i) we provide a Distribution-Free Model (DFM for short) for weighted networks. DFM has no prior distribution limitation of the adjacency matrix's elements but only an expected value related to latent structural information. DFM also allows the connectivity matrix to have negative entries. In particular, signed networks with latent community structure can be generated from DFM and the popular SBM is a sub-model of DFM.

(ii) we build a theoretical guarantee under DFM to show that a simple spectral clustering algorithm enjoys consistent estimation even when the adjacency matrix is polluted by a noise matrix. 

(iii) we summarize a four-step data generation process (Steps (a)-(d) in this paper) to model real-world weighted networks with missing edges by a combination of DFM, noise matrix, and a model for unweighted networks.

(iv) we present a performance analysis of some traditional community detection methods proposed by \cite{rohe2011spectral,RSC,SCORE,lei2015consistency, chen2018convexified} in our experimental studies and find that these methods can successfully detect communities for adjacency matrices generated by our data generation processes.

\textbf{\textit{Notations.}}
We take the following general notations in this paper. For any positive integer $m$, let $[m]:= \{1,2,\ldots,m\}$. For a vector $x$, $\|x\|_{q}$ denotes its $l_{q}$-norm. $M'$ is the transpose of the matrix $M$, and $\|M\|$ denotes the spectral norm, $\|M\|_{F}$ denotes the Frobenius norm, and $\|M\|_{0}$ denotes the $l_{0}$ norm by counting the number of nonzero entries in $M$. Let $\sigma_{k}(M)$ and $\lambda_{k}(M)$ be the $k$-th largest singular value and its corresponding eigenvalue of matrix $M$ ordered by the magnitude. $M(i,:)$ and $M(:,j)$ denote the $i$-th row and the $j$-th column of matrix $M$, respectively. $M(S,:)$ denotes the rows in the index sets $S$ of matrix $M$. $\mathrm{rank}(M)$ denotes the rank of matrix $M$. $\mathbb{R}$ is the set of real numbers and $\mathbb{R}_{+}$ is the set of nonnegative real numbers.
\section{The Distribution-Free Model}\label{modelDFM}
Consider a undirected weighted network $\mathcal{N}=(V,E)$, where $V=\{1,2,\ldots, n\}$ is the set of nodes, and $E$ is the set of edges. Let $A\in \mathbb{R}^{n\times n}$ be the symmetric adjacency matrix of $\mathcal{N}$ such that $A(i,j)$ denotes the weight between node $i$ and node $j$ for $i,j\in[n]$, and $A$ is called adjacency matrix in this paper. It should be emphasized that $A(i,j)$ can be 0, 1, or some other finite real value in this paper. We assume
\begin{align}\label{DefinSC}
\mathrm{Network~}\mathcal{N}\mathrm{~consists~of}~K~\mathrm{perceivable~nonoverlapping~ communities~}\mathcal{C}_{1},\mathcal{C}_{2},\ldots,\mathcal{C}_{K}.
\end{align}
Let $\ell$ be an $n\times 1$ vector such that $\ell(i)=k$ if node $i$ belongs to community $k$ for $i\in[n], k\in[K]$. Let $Z\in \{0,1\}^{n\times K}$ be the membership matrix of nodes such that
\begin{align}\label{DefineSPMF}
\mathrm{rank}(Z)=K, \mathrm{and~}Z(i,\ell(i))=1, \|Z(i,:)\|_{1}=1\qquad \mathrm{for~} i\in[n].
\end{align}
In Equation (\ref{DefineSPMF}), $\mathrm{rank}(Z)=K$ means that the set $\{i\in[n]:\ell(i)=k\}$ for any $k\in[K]$ is not the null set $\varnothing$, and $Z(i,\ell(i))=1, \|Z(i,:)\|_{1}=1$ mean that node $i$ only belongs to one of the $K$ communities for $i\in[n]$.

Let $n_{k}=\sum_{i\in[n]}Z(i,k)$ be the number of nodes belonging to community $k$ for $k\in[K]$. Set $n_{\mathrm{max}}=\mathrm{max}_{k\in[K]}n_{k}, n_{\mathrm{min}}=\mathrm{min}_{k\in[K]}n_{k}$. For $k\in[K]$, let $\mathcal{I}^{(k)}=\{i\in[n]: Z(i,k)=1\}$ be the set of nodes belonging to community $k$. For $k\in[K]$, select one node from $\mathcal{I}^{(k)}$ to construct the index set $\mathcal{I}$, i.e., $\mathcal{I}$ is the indices of nodes corresponding to $K$ nodes, one from each community. W.L.O.G., let $Z(\mathcal{I},:)=I_{K}$.

Let $P\in \mathbb{R}^{K\times K}$ be a matrix (call $P$ connectivity matrix in this paper) satisfying
\begin{align}\label{definP}
 P=P', 1\leq K_{0}:=\mathrm{rank}(P)\leq K,\mathrm{and~}\mathrm{max}_{k,l\in[K]}|P(k,l)|=1.
\end{align}
We'd emphasize that $P$ may have negative elements. Meanwhile, unless specified, throughout this article, $K_{0}$ and $K$ are assumed to be known integers. W.L.O.G., let $n_{K_{0}}$ be the $K_{0}$-th largest number among $\{n_{1}, n_{2}, \ldots, n_{K}\}$. For such definition, we have $\sigma_{K_{0}}(Z)=\sqrt{n_{K_{0}}}$. Since $\mathrm{rank}(P)=K_{0}$, $P$ has $K_{0}$ nonzero eigenvalues. The smallest nonzero singular value $\sigma_{K_{0}}(P)$ (call it separation parameter in this paper) measures the separation between communities, and we will let it into the theoretical error bound of algorithm fitting our model for further analysis.

Let $\rho$ be a positive value. Since we will let $\rho$ control the sparsity of  $\mathcal{N}$ when DFM reduces to SBM, we call $\rho$ sparsity parameter in this paper. Note that $\rho$ can be larger than 1 in this paper since $P$ may contain negative elements and $P$ may not be the probability matrix considered in most community detection literature, to name a few, \cite{SBM, DCSBM,rohe2011spectral, SCORE, lei2015consistency, chen2018convexified}. $\rho$'s range can be different for different distribution $\mathcal{F}$, see Examples \ref{Bernoulli}-\ref{Signed} for detail.

For arbitrary distribution $\mathcal{F}$ and all pairs of $(i,j)$ with $i,j\in[n]$, our model assumes that $A(i,j)$ are independent random variables generated according to $\mathcal{F}$ satisfying
\begin{align}\label{DefinMM}
\mathbb{E}[A(i,j)]=\Omega(i,j), \mathrm{where~}\Omega:=\rho ZPZ'.
\end{align}
Equation (\ref{DefinMM}) means that we only assume all elements of $A$ are independent random variables and $\mathbb{E}[A]=\rho ZPZ'$ without any prior knowledge on specific distribution of $A(i,j)$ for $i,j\in[n]$ since the distribution $\mathcal{F}$ can be arbitrary. Call $\Omega$ population adjacency matrix in this paper. Now, we are ready to present our model.
\begin{Definition}
Call model (\ref{DefinSC})-(\ref{DefinMM}) the Distribution-Free Model (DFM), and denote it by $DFM_{n}(K,P,Z,\rho,\mathcal{F})$.
\end{Definition}
In the naming of our model DFM, the term ``Distribution-Free'' comes from the fact that $\mathcal{F}$ can be arbitrary distribution as long as $A(i,j)$ is a random variable generated from $\mathcal{F}$ with expectation $\Omega(i,j)$. Due to the arbitrariness of $\mathcal{F}$, examples of some classical distributions such as Bernoulli, Normal, Binomial, Poisson, Exponential, and Uniform are discussed in Examples \ref{Bernoulli}-\ref{Uniform}.

The next proposition guarantees the identifiability of DFM, suggesting that DFM is well-defined.
\begin{Proposition}\label{idDFM}
(Identifiability of DFM). DFM is identifiable: For eligible  $(P,Z)$ and $(\tilde{P}, \tilde{Z})$, if $\rho ZPZ'=\rho \tilde{Z}\tilde{P}\tilde{Z}'$, then $Z=\tilde{Z}$ and $P=\tilde{P}$.
\end{Proposition}
Since the arbitrariness of $\mathcal{F}$ allows $P$ to have negative elements in Equation (\ref{definP}), $\Omega$ may have negative elements, and this suggests the generality of DFM, especially when we compare our DFM with some previous models. The details of the comparisons are given below.
\begin{itemize}
  \item Compared with the classical Stochastic Blockmodels (SBM) of \cite{SBM}, as emphasized in Equations (\ref{definP}) and (\ref{DefinMM}), P can have negative elements and there is no constraint on the distribution of $A(i,j)$ for all nodes under DFM, while $P$ must have nonnegative entries and $A(i,j)\sim\mathrm{Bernorlli}(\Omega(i,j))$ under SBM, which suggests that DFM is more applicable than SBM since DFM can capture the latent structure of more general network $\mathcal{N}$ than SBM. Sure, when $P$ is nonnegative and $\mathcal{F}$ is Bernoulli, our DFM reduces to SBM. 
\item Compared with the 1st Weighted Stochastic Blockmodels (WSBM) \cite{aicher2015learning}, WSBM requires elements of $A$ to be drawn from exponential family distribution while $\mathcal{F}$ can be arbitrary distribution as long as Equation (\ref{DefinMM}) holds under our DFM.
\item Compared with the 2nd WSBM proposed in \cite{ng2021weighted}, this WSBM requires that $Z$ follows from a multinomial distribution and all entries of $A$ are independent random variables from Gamma distribution, and all entries of $P$ are nonnegative. For comparison, under our DFM, $A$'s elements can be random variables from any distribution as long as Equation (\ref{DefinMM}) holds, $P$ can have negative elements, and there is no distribution constraint on $Z$ as long as Equation (\ref{DefineSPMF}) holds for the identifiability of DFM.
\item Compared with the 3rd WSBM considered in \cite{ahn2018hypergraph}, though this WSBM does not assume a specific distribution of $A$'s elements, it limits $P$ to having two positive distinct entries and $A$'s elements belonging to $[0,1]$. For comparison, our DFM allows $P$ to have negative elements, and our DFM also allows $A$'s elements to be negative.
\item Compared with the 4th WSBM proposed in \cite{palowitch2018significance}, though this WSBM also has no constraint on distribution $\mathcal{F}$, it requires that all entries of $P$ should be nonnegative and $\mathcal{F}$ is a distribution on the non-negative real line while our DFM allows $\mathcal{F}$ to be a distribution on the real line and $P$ to have negative elements.
\item Compared with the 5th WSBM (also known as a homogeneous weighted stochastic block model, we call it hWSBM for short) proposed in \cite{jog2015information}, though edges can also be generated from arbitrary distributions, hWSBM only models the case that $P$ has two different positive elements, which is the main limitation of hWSBM compared with our DFM. Meanwhile, another hWSBM proposed in \cite{xu2020optimal} also limits $P$ to having two different positive entries.
\end{itemize}
\begin{Remark}
The fact that DFM allows $A$ to have negative elements is not essential because we can always make $A$'s elements positive by adding a sufficiently large constant. We may emphasize that our DFM allows $A$ to have negative elements occasionally because this indicates that the connectivity matrix $P$ can have negative elements or $\mathcal{F}$ can be distributions that can generate negative values under our DFM. For example, when $\mathcal{F}$ is Normal distribution or $A$ is the adjacency matrix of a signed network, all elements of $P$ can be negative and $A$ always have negative elements. For detail, see Examples \ref{Normal}, \ref{Uniform}, and \ref{Signed}.
\end{Remark}
In the above comparisons, we emphasize the significance of the fact that our DFM allows $P$ to have negative elements while these WSBMs limit $P$ to having nonnegative entries since the mean of Normal distribution can be negative suggested by the multi-way blockmodels of \cite{airoldi2013multi} which allows all elements of the adjacency matrix to follow Normal distribution. Therefore, the above weighted models can not model the weighted network in which $A$'s elements follow a Normal distribution with a negative mean due to their nonnegative elements requirement of $P$. For comparison, our DFM can model such weighted networks, and this guarantees the generality of our DFM compared with these WSBMs.

After introducing our model DFM, we are ready to apply the following processes to generate a random adjacency matrix $A$ with true node label vector $\ell$ from the DFM:
\begin{itemize}
  \item[Step (a)] Set $\Omega=\rho ZPZ'$, where $n, K, Z, P$ should satisfy Equations (\ref{DefineSPMF})-(\ref{definP}), and whether $P$'s elements can be negative or not (and $\rho$'s range) for different distribution $\mathcal{F}$ is discussed in Examples \ref{Bernoulli}-\ref{Signed}.
  \item[Step (b)] For $1\leq i\leq j\leq n$, let $A(i,j)$ be a random variable generated from distribution $\mathcal{F}$ with expectation $\Omega(i,j)$. Set $A(j,i)=A(i,j)$ since we only consider un-directed weighted networks. If we do not consider self-connected nodes, let $A$'s diagonal entries be zeros.
\end{itemize}
When $A$ is generated by Steps (a)-(b) under the model DFM, we call this process $A\leftarrow$(DFM) for convenience.

After generating $A$ by $A\leftarrow$(DFM), in this article, we aim at answering the following questions:
\begin{itemize}
  \item[Q (1)] Can we develop a method to estimate $\ell$ with known adjacency matrix $A$ and the number of communities $K$ when $A$ is generated from arbitrary distribution $\mathcal{F}$ satisfying Equation (\ref{DefinMM}) under our DFM?
  \item[Q (2)] When there is a method such that Q (1) is solved, let $\hat{\ell}$ be the estimated node label vector. Under our DFM, can we bound the difference between $\hat{\ell}$ and $\ell$?
  \item[Q (3)] How do traditional methods designed for unweighted networks generated from SBM and DCSBM perform when $A$ is the adjacency matrix of a weighted network generated by $A\leftarrow$(DFM)?
\end{itemize}
From now on, we will answer these questions in turn.
\section{Method: DFA}
After introducing the model DFM and the process $A\leftarrow$(DFM), to answer Q (1) and Q (2), we aim at designing an algorithm to fit DFM and building a theoretical guarantee of consistent estimation for this algorithm. Designing algorithms by using the idea of likelihood maximization of \cite{SBM, DCSBM,zhao2012consistency} or pseudo-log-likelihood of \cite{amini2013pseudo} to fit DFM is inappropriate since expectation-maximization requires prior knowledge of a specific distribution of $A$'s elements while our DFM is distribution-free. Instead, designing an algorithm based on the idea of spectral clustering is a good choice to fit DFM without losing DFM's distribution-free property. In this paper, one simple spectral clustering algorithm is proposed to fit our DFM, and we introduce our algorithm from the oracle case given $\Omega$ to the real case given $A$. Since $\mathrm{rank}(Z)=K, \mathrm{rank}(P)=K_{0}$, and $K_{0}\leq K\ll n$, we have $\mathrm{rank}(\Omega)=K_{0}$ by basic algebra, i.e., $\Omega$ has a low-dimensional structure with $K_{0}$ nonzero eigenvalue values. Let $\Omega=U\Lambda U'$ be the compact eigenvalue decomposition of $\Omega$, where  $U\in\mathbb{R}^{n\times K_{0}}, \Lambda\in\mathbb{R}^{K_{0}\times K_{0}}$, and $U'U=I_{K_{0}}$ where $I_{K_{0}}$ is a $K_{0}\times K_{0}$ identity matrix.  The below lemma presents our innovation on designing an efficient approach to fit DFM from the oracle case.
\begin{Lemma}\label{GenUrUc}
Under $DFM_{n}(K,P,Z,\rho,\mathcal{F})$, we have $U=ZB$ where $B=U(\mathcal{I},:)$.
\end{Lemma}
Since $U=ZU(\mathcal{I},:)$ by Lemma \ref{GenUrUc}, we have $U(i,:)=U(j,:)$ if $\ell(i)=\ell(j)$ for $i,j\in[n]$, i.e., $U$ has $K$ distinct rows and applying k-means algorithm on all rows of $U$ assuming there are $K$ communities exactly returns nodes memberships up to a label permutation. Meanwhile, since k-means puts $i,j$ into the same community if $U(i,:)=U(j,:)$, without confusion, we still say that applying the k-means algorithm on $U$ with $K$ clusters exactly recovers $\ell$.

We are now ready to give an ideal algorithm which we call Ideal DFA. Input: $\Omega, K_{0}$ and $K$. Output: $\ell$.
\begin{itemize}
  \item Let $\Omega=U\Lambda U'$ be the top-$K_{0}$ eigendecomposition of $\Omega$ such that $U\in\mathbb{R}^{n\times K_{0}}, \Lambda\in\mathbb{R}^{K_{0}\times K_{0}},U'U=I_{K_{0}}$.
\item  Apply k-means algorithm on $U$  with $K$ communities, i.e., find a solution to the following problem
              \begin{align*}
              M^{*}=\mathrm{argmin}_{M\in \mathcal{M}_{n, K_{0}, K}}\|M-U\|^{2}_{F},
              \end{align*}
              where $\mathcal{M}_{n, K_{0}, K}$ is the set of $n\times K_{0}$ matrices with only $K$ different rows.
\item Use $M^{*}$ to obtain the labels vector $\ell$, i.e., if $M^{*}(i,:)=M^{*}(j,:)$, then nodes $i$ and $j$ are in the same community for $i,j\in[n]$.
\end{itemize}
For convenience, call the last two steps ``Apply k-means on all rows of $U$ with $K$ communities to obtain $\ell$''. By Lemma \ref{GenUrUc}, Ideal DFA exactly returns $\ell$, and this supports the identifiability of DFM in turn. To extend the ideal case to the real case, we introduce some notations for further study.

Let $W$ be an $n\times n$ symmetric random matrix such that all elements of $W$ are independent random variables satisfying
\begin{align}\label{DefinW}
\mathbb{E}[W]=0_{n\times n} \mathrm{~and~}\sigma^{2}_{W}=\mathrm{max}_{i,j\in[n]}\mathrm{Var}(W(i,j))\mathrm{~is~finite},
\end{align}
where $0_{n\times n}$ denotes the $n\times n$ matrix with all elements being zeros, $\mathrm{Var}(W(i,j))$ denotes the variance of $W(i,j)$, and we let $\sigma^{2}_{W}$ be the maximum variance of $W$'s entries without confusion with the singular value notation. Note that there is no distribution constraint of $W$'s elements.  Call the $W$ noise matrix in this paper.

Since $A(i,j)$ can be any finite real numbers for $i,j\in[n]$ under DFM, we let $\hat{A}=A+W$ and call $\hat{A}$ the observed adjacency matrix, i.e., we consider the following step:
\begin{itemize}
  \item[Step (c)] Set $\hat{A}=A+W$, where $A$ is obtained by $A\leftarrow$(DFM).
\end{itemize}
For convenience, we call Step (c) $\hat{A}\leftarrow$(DFM+noises).  We introduce the noise matrix $W$ mainly for the fact that some entries of $A$ may be slightly perturbed by noise. Since $A$'s elements are finite real numbers, such noise may occur from measurement error when recording the elements of $A$. Sure, if $W=0_{n\times n}$ which suggests that there is no noise when recording $A$, the observed adjacency matrix $\hat{A}$ equals to the adjacency matrix $A$. Instead, we assume that there exists some noise such that $W$ may not be a zero matrix, and how the noise matrix $W$ influences the theoretical and numerical studies of the method proposed to fit our model will be studied in this article.

The community labels $\ell$ are unknown, and we aim at using $(\hat{A}, K)$ to predict them when $A$ is generated from DFM and $W$ is generated satisfying Equation (\ref{DefinW}). Let $\tilde{A}=\hat{U}\hat{\Lambda}\hat{U}'$ be the top-$K_{0}$-dimensional eigendecomposition of the observed adjacency matrix $\hat{A}$ such that $\hat{U}\in \mathbb{R}^{n\times K_{0}}, \hat{\Lambda}\in \mathbb{R}^{K_{0}\times K_{0}}, \hat{U}'\hat{U}=I_{K_{0}}$, and $\hat{\Lambda}$ contains the top $K_{0}$ eigenvalues of $\hat{A}$. Algorithm \ref{alg:DFA} called DFA is a natural extension of the Ideal DFA to the real case. We use DFA to name this algorithm to emphasize its distribution-free property since it is designed to fit our DFM. When $P$ is full rank such that $K_{0}=K$ under DFM and the noise matrix $W$ is a zero matrix such that the observed adjacency matrix $\hat{A}$ equals to the adjacency matrix $A$, DFA is the oPCA algorithm mentioned in \cite{SCORE}, and the Algorithm 1 of \cite{lei2015consistency} using k-means to replace their $(1+\epsilon)$ approximate k-means algorithm. 
\begin{algorithm}
\caption{\textbf{Distribution-Free Algorithm (DFA)}}
\label{alg:DFA}
\begin{algorithmic}[1]
\Require The observed adjacency matrix $\hat{A}\in \mathbb{R}^{n\times n}$, number of communities $K$, and $P$'s rank $K_{0}$.
\Ensure The estimated $n\times 1$ labels vector $\hat{\ell}$.
\State Let $\tilde{A}=\hat{U}\hat{\Lambda}\hat{U}'$ be the top-$K_{0}$ eigendecomposition of $\hat{A}$ such that $\hat{\Lambda}$ contains the leading $K_{0}$ eigenvalues of $\hat{A}$ and $\hat{U}'\hat{U}=I_{K_{0}}$.
\State Apply k-means on all rows of $\hat{U}$ with $K$ clusters to obtain $\hat{\ell}$.
\end{algorithmic}
\end{algorithm}
\begin{Remark}
The goal of DFA is to estimate $\ell$ from the observed adjacency matrix $\hat{A}$ under DFM. Following similar idea of steps 8 and 9 in Algorithm 1 of \cite{SPACL}, we can also recover $\rho$ and $P$ from $\hat{A}$ based on DFA. We introduce the idea from the oracle case with known $\Omega$ to the real case with given $\hat{A}$.
\begin{itemize}
  \item Oracle case: By the facts that $\Omega=Z \rho PZ'$ and $\Omega=U\Lambda U'$, we have $\Omega(\mathcal{I},\mathcal{I})=U(\mathcal{I},:)\Lambda U(\mathcal{I},:)'=Z(\mathcal{I},:)\rho PZ(\mathcal{I},:)'=\rho P$, i.e., $\rho P=U(\mathcal{I},:)\Lambda U(\mathcal{I},:)'$.
  \item Real case:  After obtaining $\hat{\ell}$ by DFA, let $\hat{\mathcal{I}}^{(k)}=\{i\in[n]: \hat{\ell}(i)=k\}$ be the set of nodes belonging to estimated community $k$ for $k\in[K]$. Select one node from $\hat{\mathcal{I}}^{(k)}$ to construct the estimated index set $\hat{\mathcal{I}}$, i.e., $\hat{\mathcal{I}}$ is a estimation of $\mathcal{I}$. By the oracle case, we see that $\hat{U}(\hat{\mathcal{I}},:)\hat{\Lambda}\hat{U}(\hat{\mathcal{I}},:)'$ should be a good estimation of $\rho P$. Let $\hat{\rho}$ be the maximum entry of $|\hat{U}(\hat{\mathcal{I}},:)\hat{\Lambda}\hat{U}(\hat{\mathcal{I}},:)'|$ and $\hat{P}=\frac{\hat{U}(\hat{\mathcal{I}},:)\hat{\Lambda}\hat{U}(\hat{\mathcal{I}},:)'}{\hat{\rho}}$, where $\hat{\rho}$ and $\hat{P}$ are estimations of $\rho$ and the connectivity matrix $P$, respectively.
\end{itemize}
\end{Remark}

Next, we aim to study DFA's consistency under DFM. For convenience, set $\tau=\mathrm{max}_{i,j\in[n]}|A(i,j)+W(i,j)-\Omega(i,j)|$ and $\gamma=\frac{\sigma^{2}_{A}}{\rho}$, where $\sigma^{2}_{A}=\mathrm{max}_{i,j\in[n]}\mathrm{Var}(A(i,j))$. Note that if we consider a specific distribution, $\tau$ and $\gamma$ are directly related to this distribution, see Examples \ref{Bernoulli}-\ref{Signed} for detail. We also need the following assumption for a theoretical guarantee of DFA's performance.
\begin{Assumption}\label{assumesparsity}
Assume that $\gamma\rho n+\sigma^{2}_{W}n\geq\tau^{2}\mathrm{log}(n)$.
\end{Assumption}
Assumption \ref{assumesparsity} provides a lower bound requirement of $\gamma\rho n$ for our theoretical analysis. Since $\gamma$ may depend on $\rho$ and it may be different for different distribution $\mathcal{F}$, we can obtain the exact form of Assumption \ref{assumesparsity} for a specific distribution. For detail, see Examples \ref{Bernoulli}-\ref{Signed}. Now we are ready to bound $\|\hat{A}-\Omega\|$ as below.
\begin{Lemma}\label{BoundAOmega}
Under $DFM_{n}(K,P,Z,\rho,\mathcal{F})$, when Assumption \ref{assumesparsity} holds, with probability at least $1-o(n^{-\alpha})$ for any $\alpha>0$, we have
\begin{align*}
\|\hat{A}-\Omega\|=O(\sqrt{(\gamma\rho n+\sigma^{2}_{W}n)\mathrm{log}(n)}).
\end{align*}
\end{Lemma}
In the proof of Lemma \ref{BoundAOmega}, there is no requirement on $A$'s diagonal elements, so DFA can also detect communities when there exist self-connected nodes in the weighted network $\mathcal{N}$.

We consider the performance criterion defined in \cite{joseph2016impact} to measure the estimation error of DFA. This measurement is given below. Let $\{\hat{\mathcal{C}}_{1}, \hat{\mathcal{C}}_{2}, \ldots, \hat{\mathcal{C}}_{K}\}$ be the estimated partition of nodes $\{1,2,\ldots, n\}$ obtained from $\hat{\ell}$ in algorithm \ref{alg:DFA} such that $\hat{\mathcal{C}}_{k}=\{i: \hat{\ell}(i)=k\}$ for $k\in[K]$. Define the criterion as
\begin{align*}
\hat{f}=\mathrm{min}_{\pi\in S_{K}}\mathrm{max}_{k\in[K]}\frac{|\mathcal{C}_{k}\cap \hat{\mathcal{C}}^{c}_{\pi(k)}|+|\mathcal{C}^{c}_{k}\cap \hat{\mathcal{C}}_{\pi(k)}|}{n_{k}},
\end{align*}
where $S_{K}$ is the set of all permutations of $\{1,2,\ldots, K\}$ and the superscript $c$ denotes a complementary set. As mentioned in \cite{joseph2016impact}, $\hat{f}$ measures the maximum proportion of nodes in the symmetric difference of $\mathcal{C}_{k}$ and $\hat{\mathcal{C}}_{\pi(k)}$. The theoretical upper bound of $\hat{f}$ is provided by the following theorem, which guarantees the estimation consistency of DFA under DFM.
\begin{Theorem}\label{MainDFA}
Under $DFM_{n}(K,P,Z,\rho,\mathcal{F})$, let $\hat{\ell}$ be obtained from Algorithm \ref{alg:DFA}, when Assumption \ref{assumesparsity} holds, with probability at least $1-o(n^{-\alpha})$, we have
\begin{align*}
\hat{f}=O(\frac{K_{0}K(\gamma\rho n+\sigma^{2}_{W}n)\mathrm{log}(n)}{\sigma^{2}_{K_{0}}(P)\rho^{2}\delta^{2}n^{2}_{K_{0}}n_{\mathrm{min}}}),
\end{align*}
where $\delta=\mathrm{min}_{k\neq l}\|B(k,:)-B(l,:)\|_{F}$. Furthermore, we have below three special cases.
\begin{itemize}
\item Case (I): when $K_{0}=K$, we have
\begin{align*}
\hat{f}=O(\frac{K^{2}(\gamma\rho n+\sigma^{2}_{W}n)n_{\mathrm{max}}\mathrm{log}(n)}{\sigma^{2}_{K}(P)\rho^{2}n^{3}_{\mathrm{min}}}).
\end{align*}
\item Case (II): when $K_{0}=K=O(1)$ and $\frac{n_{\mathrm{max}}}{n_{\mathrm{min}}}=O(1)$, we have
\begin{align*}
\hat{f}=O(\frac{(\gamma\rho n+\sigma^{2}_{W}n)\mathrm{log}(n)}{\sigma^{2}_{K}(P)\rho^{2}n^{2}}).
\end{align*}
\item Case (III): when $K_{0}=K=O(1), \frac{n_{\mathrm{max}}}{n_{\mathrm{min}}}=O(1)$ and $\sigma^{2}_{W}=0$ (i.e., the case when $W$ is a zero matrix), we have
\begin{align*}
\hat{f}=O(\frac{\gamma\mathrm{log}(n)}{\sigma^{2}_{K}(P)\rho n}).
\end{align*}
\end{itemize}
\end{Theorem}
Since DFM has no distribution constraint on $A$ as long as $\mathbb{E}[A]=\Omega$ under any distribution $\mathcal{F}$, Theorem \ref{MainDFA} provides a general upper bound on DFA's estimation error. From Theorem \ref{MainDFA}, we find that by increasing the variance of the noise matrix $W$, the error rate of DFA increases, and this phenomenon is consistent with our intuition that if the observed matrix $\hat{A}$ differs with the adjacency matrix $A$ too much caused by the noise matrix, then $\hat{A}$ deviates from $\Omega$ a lot since $A$ is generated under DFM. Theorem \ref{MainDFA} also says that increasing $\sigma^{2}_{K}(P)$ decreases DFA's error rate. Especially, for Case (III), Theorem \ref{MainDFA} says that $\sigma_{K}(P)$ should shrink slower than $\sqrt{\frac{\gamma\mathrm{log}(n)}{\rho n}}$ to make DFA's error rate sufficiently small with high probability. For the influence of $\rho$ on DFA's performance, we need to consider the property for a specific distribution since $\gamma$ may be related to $\rho$, and we analyze $\rho$'s influence on DFA's performance in Examples \ref{Bernoulli}-\ref{Signed}.

From now on, we only consider Case (III) for the convenience of our further theoretical analysis. If $A$ is further assumed to follow a specific distribution such that we can obtain $\gamma$'s upper bound, then the exact theoretical upper bound of DFA's error rate can be obtained immediately from Theorem \ref{MainDFA}. For $i,j\in[n]$, we provide some examples to support our statement that results in Theorem \ref{MainDFA} are general:
\begin{Example}\label{Bernoulli}
When $\mathcal{F}$ is \textbf{Bernoulli distribution} such that $A(i,j)\sim\mathrm{Bernorlli}(\Omega(i,j))$, i.e., $A\in\{0,1\}^{n\times n}$. Under Bernoulli distribution, $\rho P$ is a probability matrix, so $P$'s elements should be nonnegative and $\rho$'s range is $(0,1]$ because $\mathrm{max}_{k,l\in[K]}P(k,l)=1$ by Equation (\ref{definP}). Since $\mathrm{Var}(A(i,j))=\Omega(i,j)(1-\Omega(i,j))\leq \Omega(i,j)\leq \rho$ when  $A(i,j)\sim\mathrm{Bernorlli}(\Omega(i,j))$, we have $\gamma=1$. For $\tau$, it is 1 for this case. Then, Assumption \ref{assumesparsity} becomes $\rho n\geq \mathrm{log}(n)$, and it provides a lower bound requirement on $\rho$. The theoretical upper bound of DFA's estimation error obtained immediately from Case (III) is $\hat{f}=O(\frac{\mathrm{log}(n)}{\sigma^{2}_{K}(P)\rho n})$, and it matches Corollary 3.2 of \cite{lei2015consistency} up to logarithmic factor which suggests the optimality of our theoretical results. From the theoretical upper bound of DFA's error rate, we see that increasing $\rho$ decreases DFA's error rate. So, $\rho$ controls network sparsity for Bernoulli distribution. Furthermore, by the separation condition and sharp threshold criterion developed in \cite{Criterion}, we know that the sparsity requirement of Assumption \ref{assumesparsity} and  $\hat{f}=O(\frac{\mathrm{log}(n)}{\sigma^{2}_{K}(P)\rho n})$ are theoretical optimal.
\end{Example}
\begin{Example}\label{Normal}
When $\mathcal{F}$ is \textbf{Normal distribution} such that  $A(i,j)\sim \mathrm{Normal}(\Omega(i,j),\sigma^{2}_{A})$, i.e., $A\in\mathbb{R}^{n\times n}$. Under Normal distribution, $P$ can have negative elements under DFM, $\rho$'s range is $(0,+\infty)$ because $\rho P$ is not a probability matrix, $\tau$ is an unknown finite number, and $\gamma=\frac{\sigma^{2}_{A}}{\rho}$. Setting $\gamma=\frac{\sigma^{2}_{A}}{\rho}$, Assumption \ref{assumesparsity} becomes $\sigma^{2}_{A}n\geq\tau^{2}\mathrm{log}(n)$, and it does not mean a lower bound requirement on $\rho$ but a lower bound requirement on network size $n$ for our theoretical analysis. Setting $\gamma=\frac{\sigma^{2}_{A}}{\rho}$, for Case (III), $\hat{f}=O(\frac{\sigma^{2}_{A}\mathrm{log}(n)}{\sigma^{2}_{K}(P)\rho^{2} n})$. Therefore, for Normal distribution, increasing $\rho$ (or decreasing $\sigma^{2}_{A}$) decreases DFA's error rate. An extreme case is when $\sigma^{2}_{A}=0$, we have  $A(i,j)=\Omega(i,j)$ for $i,j\in[n]$, and for such case, the error rate is 0 surely.
\end{Example}
\begin{Example}\label{Binomial}
When $\mathcal{F}$ is \textbf{Binomial distribution} such that  $A(i,j)\sim \mathrm{Binomial}(m,\frac{\Omega(i,j)}{m})$ for some positive integer $m$, i.e., $A\in\{0,1,2,\ldots,m\}^{n\times n}$. Under Binomial distribution, we have $\mathbb{E}[A(i,j)]=\Omega(i,j)$ satisfying Equation (\ref{DefinMM}) under DFM, $\frac{\Omega(i,j)}{m}$ is a probability suggesting $P$ should have nonnegative elements and $\rho$'s range is $(0,m]$, $\tau=m$, and $\gamma=\mathrm{max}_{i,j\in[n]}m\frac{\Omega(i,j)}{m}(1-\frac{\Omega(i,j)}{m})/\rho\leq \mathrm{max}_{i,j\in[n]}\frac{\Omega(i,j)}{\rho}\leq 1$, i.e., $\gamma$ is finite. Setting $\tau=m, \gamma=1$, Assumption \ref{assumesparsity} becomes $\rho n\geq m^{2}\mathrm{log}(n)$, which means a lower bound requirement on $\rho$. Setting $\gamma=1$, for Case (III), $\hat{f}=O(\frac{\mathrm{log}(n)}{\sigma^{2}_{K}(P)\rho n})$. Therefore, for Binomial distribution, increasing $\rho$ decreases DFA's error rate.
\end{Example}
\begin{Example}\label{Poisson}
When $\mathcal{F}$ is \textbf{Poisson distribution} such that  $A(i,j)\sim \mathrm{Poisson}(\Omega(i,j))$, i.e., $A\in\mathbb{Z}^{n\times n}$, where $\mathbb{Z}$ is the set of non-negative integers. By the property of Poisson distribution, we have $\mathbb{E}[A(i,j)]=\Omega(i,j)$ satisfying Equation (\ref{DefinMM}), all entries of $P$ should be nonnegative under DFM, $\rho$'s range is $(0,+\infty)$ because $\rho P$ is not a probability matrix, $\tau$ is an unknown finite integer, and $\gamma=\mathrm{max}_{i,j\in[n]}\frac{\mathrm{Var}(A(i,j))}{\rho}=\mathrm{max}_{i,j\in[n]}\frac{\Omega(i,j)}{\rho}\leq 1$. Setting $\gamma=1$, Assumption \ref{assumesparsity} becomes $\rho n\geq\tau^{2}\mathrm{log}(n)$ which provides a lower bound requirement on $\rho$. Setting $\gamma=1$, for Case (III), $\hat{f}=O(\frac{\mathrm{log}(n)}{\sigma^{2}_{K}(P)\rho n})$. Therefore, increasing $\rho$ decreases DFA's error rate for Poisson distribution.
\end{Example}
\begin{Example}\label{Exponential}
When $\mathcal{F}$ is \textbf{Exponential distribution} such that  $A(i,j)\sim\mathrm{Exponential}(\frac{1}{\Omega(i,j)})$, i.e, $A\in\mathbb{R}_{+}^{n\times n}$. For Exponential distribution, all elements of $P$ should be larger than $0$ to make the Exponential parameter $\Omega(i,j)$ well-defined under DFM. For Exponential distribution, $\rho$'s range is $(0,+\infty)$ because $\rho P$ is not a probability matrix. Since $\mathbb{E}[A(i,j)]=\Omega(i,j)$, Equation (\ref{DefinMM}) is satisfied. For Exponential distribution, $\tau$ is an unknown finite number and $\gamma=\mathrm{max}_{i,j\in[n]}\mathrm{Var}(A(i,j))/\rho=\mathrm{max}_{i,j\in[n]}\Omega^{2}(i,j)/\rho\leq \rho$. Setting $\gamma=\rho$,  Assumption \ref{assumesparsity} is $\rho^{2}n\geq\tau^{2}\mathrm{log}(n)$, and it means a lower bound requirement on $\rho$. Setting $\gamma=\rho$, for Case (III), $\hat{f}=O(\frac{\mathrm{log}(n)}{\sigma^{2}_{K}(P)n})$. So, increasing $\rho$ has no significant influence on DFA's performance and DFA's error rate is always quite small as long as Assumption \ref{assumesparsity} holds for $\rho$ and $n$ is a large number for Exponential distribution to make the probability $1-o(n^{-\alpha})$ close to 1.
\end{Example}
\begin{Example}\label{Uniform}
When $\mathcal{F}$ is \textbf{Uniform distribution}, there are three cases.
\begin{itemize}
  \item Uniform-Case (1): $A(i,j)\sim\mathrm{Uniform}[0,2\Omega(i,j)]$, i.e., $A(i,j)\in[0,2\rho]$. For this case, all elements of $P$ should be nonnegative, $\rho$'s range is $(0,+\infty)$ because $\rho P$ is not a probability matrix, $\tau$ is an unknown finite value, $\mathbb{E}[A(i,j)]=\frac{0+2\Omega(i,j)}{2}=\Omega(i,j)$ satisfying Eq (\ref{DefinMM}), and $\mathrm{Var}(A(i,j))=\frac{4\Omega^{2}(i,j)}{12}\leq \frac{\rho^{2}}{3}$, i.e., $\gamma\leq\frac{\rho}{3}$ is finite. Setting $\gamma=\frac{\rho}{3}$, Assumption \ref{assumesparsity} is $\rho^{2}n\geq3\tau^{2}\mathrm{log}(n)$ and it means a lower bound requirement of $\rho$. Setting $\gamma=\frac{\rho}{3}$, for Case (III), $\hat{f}=O(\frac{\mathrm{log}(n)}{\sigma^{2}_{K}(P)n})$.  So, increasing $\rho$ does not influence DFA's performance for Uniform-Case (1).
  \item Uniform-Case (2): $A(i,j)\sim\mathrm{Uniform}[2\Omega(i,j),0]$, i.e., $A(i,j)\in[-2\rho,0]$. For this case, all elements of $P$ should be non-positive, $\rho$'s range is also $(0,+\infty)$, $\tau$ is also an unknown finite number, $\mathbb{E}[A(i,j)]=\Omega(i,j)$ satisfying Eq (\ref{DefinMM}), and $\mathrm{Var}(A(i,j))=\frac{4\Omega^{2}(i,j)}{12}\leq \frac{\rho^{2}}{3}$, i.e., $\gamma\leq\frac{\rho}{3}$. Setting $\gamma=\frac{\rho}{3}$, Assumption \ref{assumesparsity} is also $\rho^{2}n\geq3\tau^{2}\mathrm{log}(n)$. Setting $\gamma=\frac{\rho}{3}$, for Case (III), $\hat{f}=O(\frac{\mathrm{log}(n)}{\sigma^{2}_{K}(P)n})$ which is the same as Uniform-Case (1).
  \item Uniform-Case (3): $A(i,j)\sim\mathrm{Unifrom}(u,2\Omega(i,j)-u)$ or $A(i,j)\sim\mathrm{Unifrom}(2\Omega(i,j)-u,u)$ for $\mu\in\mathbb{R}$ as long as Eq (\ref{DefinMM}) holds when $\mathcal{F}$ is Uniform distribution. The analysis is similar to Uniform-Case (I) and Uniform-Case (II), and we omit it here.
\end{itemize}
\end{Example}
\begin{Example}\label{Signed}
DFM can also model \textbf{signed networks} by setting $\mathbb{P}(A(i,j)=1)=\frac{1+\Omega(i,j)}{2}$ and $\mathbb{P}(A(i,j)=-1)=\frac{1-\Omega(i,j)}{2}$, i.e., $A\in\{-1,1\}^{n\times n}$. For signed networks, all elements of $P$ are real values, $\rho$'s range is  $(0,1]$ since $\frac{1+\Omega(i,j)}{2}$ and $\frac{1-\Omega(i,j)}{2}$ are probabilities, $\mathbb{E}[A(i,j)]=\Omega(i,j)$ satisfying Equation (\ref{DefinMM}), and $\mathrm{Var}(A(i,j))=1-\Omega^{2}(i,j)\leq1$, i.e., $\gamma\leq \frac{1}{\rho}$. For $\tau$, it is less than 2. Setting $\tau=2, \gamma=\frac{1}{\rho}$, Assumption \ref{assumesparsity} becomes $\frac{n}{\mathrm{log}(n)}\geq 4$ which controls network size for our theoretical analysis. For Case (III), $\hat{f}=O(\frac{\mathrm{log}(n)}{\sigma^{2}_{K}(P)\rho^{2}n})$. Therefore, for signed networks, increasing $\rho$ decreases DFA's error rate.
\end{Example}
More than the above examples, DFM allows $\mathcal{F}$ to be any distribution as long as Equation (\ref{DefinMM}) holds. For example, $\mathcal{F}$ can be Double exponential, Gamma, Laplace, and Geometric distributions in \url{http://www.stat.rice.edu/~dobelman/courses/texts/distributions.c&b.pdf}. Note that Equation (\ref{DefinMM}) does not hold for some distributions. For example, the mean of t-distribution is 0 and the mean of Cauchy distribution does not exist. So $\mathcal{F}$ can not be t-distribution or Cauchy distribution.
\section{Missing edge}
From Examples \ref{Normal}, \ref{Exponential}, and \ref{Signed}, we see that there is always an edge weight between any node pair $(i,j)$ for the observed adjacency matrix $\hat{A}$ generated from the process $\hat{A}\leftarrow$(DFM+noises). However, many node pairs have no connections in real-world weighted networks. In this article, similar to \cite{xu2020optimal}, we treat an edge with weight 0 as a missing edge. There are many ways to generate missing edges in $\tilde{A}$, here we propose one general idea. Let $\mathcal{M}$ be a statistical model for un-directed unweighted networks and let $\mathcal{A}$ be an adjacency matrix for an unweighted network generated from model $\mathcal{M}$, i.e., $\mathcal{A}=\mathcal{A}'$ and $\mathcal{A}(i,j)\in\{0,1\}$ for $i,j\in[n]$. For example, when $\mathcal{M}$ is the Erd\"os-R\'enyi random graph $G(n,p)$ \cite{erdos1960evolution}, we have $\mathbb{P}(\mathcal{A}(i,j)=1)=p$ and $\mathbb{P}(\mathcal{A}(i,j)=0)=1-p$, where $p\in[0,1]$ is a probability. $\mathcal{M}$ can also be the SBM \cite{SBM}, DCSBM \cite{DCSBM} or MMSB \cite{MMSB} as long as $\mathcal{A}$ is the adjacency matrix of a unweighted network. After defining $\mathcal{A}$, now we are ready to generate missing edges in $\hat{A}$ in the following step:
\begin{itemize}
  \item[Step (d)] Obtain $\hat{A}$ by $\hat{A}\leftarrow$(DFM+noises) and $\mathcal{A}$ by model $\mathcal{M}$. Let $\hat{A}_{\mathcal{A}}$ be an $n\times n$ adjacency matrix such that its $(i,j)$-th element is $\hat{A}_{\mathcal{A}}(i,j)=\mathcal{A}(i,j)\hat{A}(i,j)$ for $i,j\in[n]$.
\end{itemize}
For convenience, we call Step (d) $\hat{A}_{\mathcal{A}}\leftarrow$(DFM+noises+$\mathcal{M}$). Without confusion, we also call $\hat{A}_{\mathcal{A}}$ adjacency matrix. Since $\mathcal{A}\in\{0,1\}^{n\times n}$, we see that there are many missing edges in $\hat{A}_{\mathcal{A}}$. Meanwhile, when $\mathcal{M}$ is the Erd\"os-R\'enyi random graph $G(n,p)$, $\hat{A}_{\mathcal{A}}$ loses more node label information of $\ell$ when $p$ decreases since a smaller $p$ indicates more missing edges in $\hat{A}_{\mathcal{A}}$. We will study the performances of DFA and some traditional community detection methods when the input adjacency matrix is $\hat{A}_{\mathcal{A}}$ and the model $\mathcal{M}$ is $G(n,p)$ in Section \ref{Experiments}.
\section{Experiments}\label{Experiments}
In this section, we conduct experimental studies of DFA's performance on simulated data and some real-world networks. In addition to DFA, four traditional methods of community detection designed under SBM or DCSBM are applied to our simulation ensemble. The first one is the spectral clustering (SC for short) algorithm studied in \cite{rohe2011spectral}, where SC is designed based on the Laplacian matrix to fit SBM. The second one is the regularized spectral clustering (RSC for short) algorithm by \cite{RSC} to fit DCSBM. The third one is the spectral clustering on ratios-of-eigenvectors (SCORE for short) algorithm by \cite{SCORE} to fit DCSBM. The last one is the convexified modularity maximization (CMM for short) method by \cite{chen2018convexified} to fit DCSBM. Unlike the theoretical studies where we allow $K_{0}\leq K$ to show that DFM is identifiable even when $K_{0}\leq K$, in the numerical study part, we always set $K_{0}=K$ since this setting is common in both numerical and empirical studies.
\subsection{Performance on synthetic networks}
\begin{figure}
\centering
\subfigure[Changing $\rho$]{\includegraphics[width=0.325\textwidth]{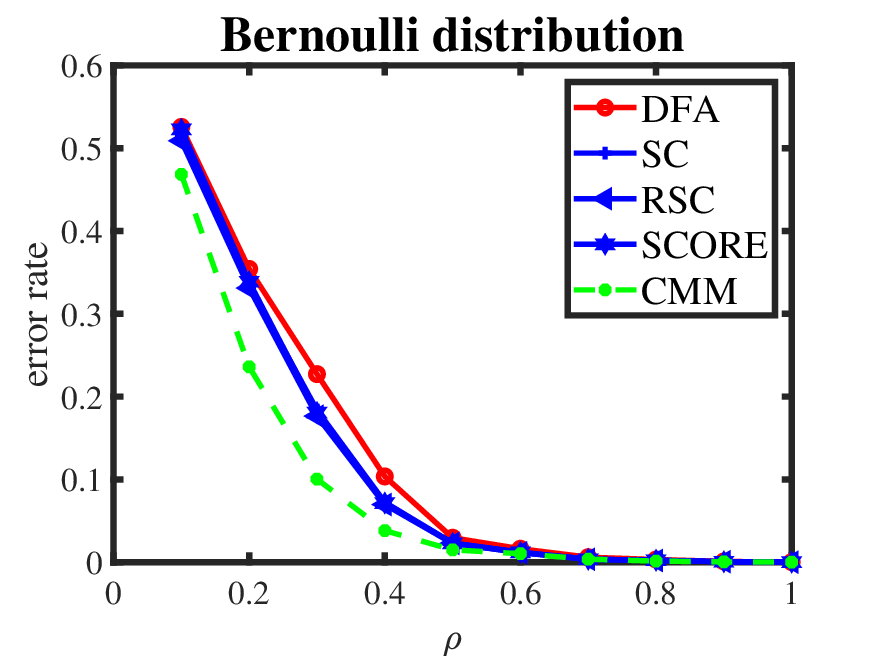}}
\subfigure[Changing $\sigma^{2}_{W}$]{\includegraphics[width=0.325\textwidth]{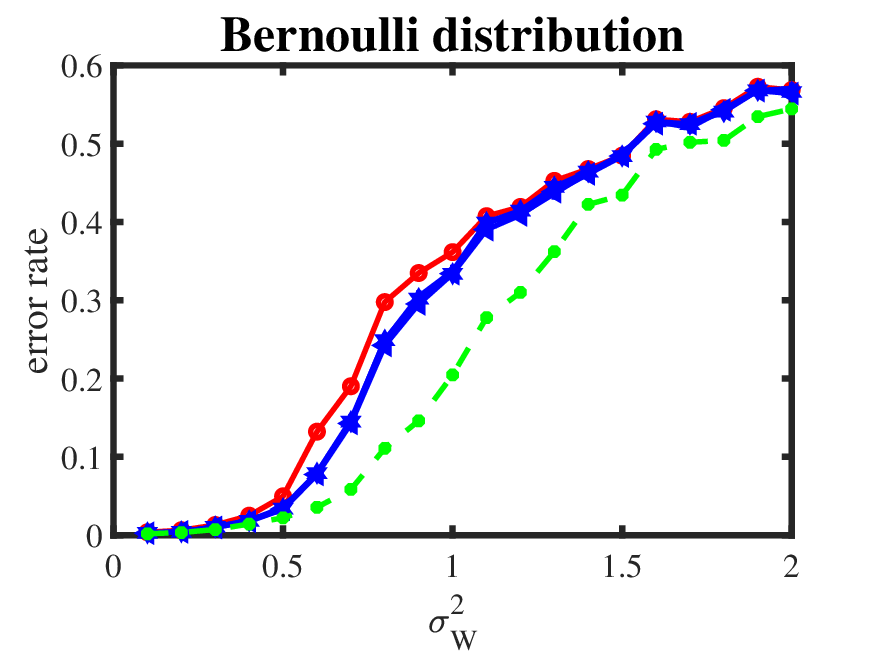}}
\subfigure[Changing $p$]{\includegraphics[width=0.325\textwidth]{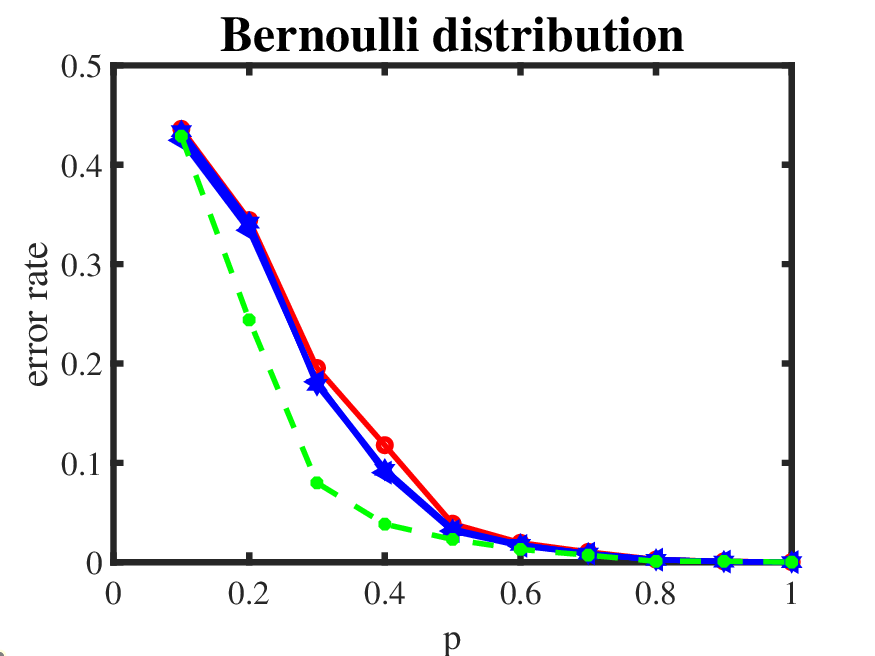}}
\caption{Numerical results of Experiment 1.}
\label{EX1} 
\end{figure}
This subsection studies the numerical performances of DFA, SC, RSC, SCORE, and CMM under different settings when $A$ follows different distributions and $\mathcal{M}$ is $G(n,p)$. Similar to the worst case relative error introduced in \cite{lei2015consistency}, $\hat{f}$ is a stronger criterion than the Hamming error introduced in \cite{SCORE} (Hamming error is same as the overall relative error of \cite{lei2015consistency}) because $\hat{f}$ may be large for some small communities as shown by Theorem \ref{MainDFA}. In the numerical study part, instead of using $\hat{f}$, we apply the Hamming error to measure the performance of each method. This error rate is defined as
\begin{align*}
  n^{-1}\mathrm{min}_{J\in\mathcal{P}_{K}}\|\hat{Z}J-Z\|_{0},
\end{align*}
where $\mathcal{P}_{K}$ is the set of all $K\times K$ permutation matrices, and the $n\times K$ matrix $\hat{Z}$ is defined as $\hat{Z}(i,\hat{\ell}(i))=1$ and all other $K-1$ entries of $\hat{Z}(i,:)$ are 0 for $i\in[n]$.

In all synthetic networks, unless specified, set $n=200, K_{0}=K=3,$ and generate $\ell$ such that each node belongs to one of the communities with equal probability. For distributions that require all elements of $P$ to be nonnegative, we set $P$ as
\[P_{1}=\begin{bmatrix}
    1&0.4&0.5\\
    0.4&0.9&0.2\\
    0.5&0.2&0.8\\
\end{bmatrix}.\]
Unless specified, for distributions that allow $P$ to have negative elements, we set $P$ as
\[P_{2}=\begin{bmatrix}
    -1&-0.4&0.5\\
    -0.4&0.9&0.2\\
    0.5&0.2&0.8\\
\end{bmatrix},\]
or $P_{3}=-P_{1}$, where all entries of $P_{3}$ are negative. For the Erd\"os-R\'enyi random graph $G(n,p)$, unless specified, we set $p=0.8$.
\begin{Remark}
For our numerical studies, there is no specific requirement on setting $P$ as long as Equation (\ref{definP}) holds and $P$'s elements should be nonnegative or positive or can be negative depending on a specific distribution $\mathcal{F}$ as analyzed in Examples \ref{Bernoulli}-\ref{Signed}. For the choosing of $\rho$, $\rho$ should be set in its range for a specific distribution as analyzed in Examples \ref{Bernoulli}-\ref{Signed}. For the setting of $Z$, there is also no specific requirement as long as Equation (\ref{DefineSPMF}) holds, i.e., each node only belongs to one community and each community has at least one node. For $G(n,p)$,  there is also no specific requirement on $p$ as long as $p\geq \frac{\mathrm{log}(n)}{n}$ such that $\mathcal{A}$ is connected with high probability \cite{abbe2017community}.
\end{Remark}
\begin{Remark}
For distributions like Normal distribution, Uniform-Case (2), and signed networks, since $\hat{A}_{\mathcal{A}}$ has negative elements for these distributions while SC and RSC are designed based on Laplacian matrix, to make SC and RSC work for adjacency matrix with negative elements, we make all elements of the adjacency matrix positive by adding a sufficiently large positive constant. For DFA, SCORE, and CMM, they always work even when the adjacency matrix has negative elements since they are designed based on the adjacency matrix instead of the Laplacian matrix.
\end{Remark}
After having $P, \rho, p$, and $Z$, to generate $\hat{A}_{\mathcal{A}}$ with $K$ communities from the distribution $\mathcal{F}$ under our DFM, each simulation experiment contains the following steps:

\begin{itemize}
  \item [Step (e)] Apply DFA (and the other four methods) to $\hat{A}_{\mathcal{A}}$ with $K$ communities, where $\hat{A}_{\mathcal{A}}$ is obtained by $\hat{A}_{\mathcal{A}}\leftarrow$(DFM+noises+$G(n,p)$). Record error rate.
  \item [Step (f)] Repeat steps (a)-(e) 50 times and report the average error rate.
\end{itemize}
We consider the following simulation setups.
\subsubsection{Experiment 1: Bernoulli distribution}
When $A(i,j)\sim \mathrm{Bernoulli}(\Omega(i,j))$ for $i,j\in[n]$, by Example \ref{Bernoulli}, we set $P$ as $P_{1}$.

\texttt{Experiment 1[a]: Changing $\rho$.} Let $W$ be a zero matrix and $\rho$ range in $\{0.1,0.2,\ldots,1\}$. Panel (a) of Figure \ref{EX1} shows the result. We see that, when the network $\mathcal{N}$ becomes denser as $\rho$ increases, DFA performs better, which is consistent with our findings in Example \ref{Bernoulli}. The other four methods perform similarly to our DFA.

\texttt{Experiment 1[b]: Changing $\sigma^{2}_{W}$.} In this sub-experiment, we study the influence of $W$ on the performances of these methods by changing $\sigma^{2}_{W}$. Set $\rho=0.8$. Let $\sigma^{2}_{W}$ range in $\{0.1, 0.2, \ldots, 2\}$, and the result is displayed by the panel (b) of Figure \ref{EX1}. We see that DFA performs poorer as $\sigma^{2}_{W}$ increases which is consistent with our theoretical finding in Theorem \ref{MainDFA}. We also see that SC, RSC, and SCORE enjoy similar performances as our DFA and these methods perform slightly poorer than CMM. The result also shows that all methods enjoy the stable performance when $A$ is slightly polluted by the noise matrix $W$, i.e., the case when $\sigma^{2}_{W}$ is small.

\texttt{Experiment 1[c]: Changing $p$.} In this sub-experiment, we study the influence of $p$ on the performances of different methods, where a smaller $p$ in $G(n,p)$ indicates more missing edges in the adjacency matrix $\hat{A}_{\mathcal{A}}$. Let $\rho=0.8$ and $W$ be a zero matrix. Let $p$ range in $\{0.1, 0.2, \ldots, 1\}$, and the result is displayed by the panel (c) of Figure \ref{EX1}. We see that all methods perform better as $p$ increases and this is consistent with our intuition that a community detection method should perform better when there are lesser missing edges in the adjacency matrix.
\subsubsection{Experiment 2: Normal distribution}
\begin{figure}
\centering
\subfigure[Changing $\rho$]{\includegraphics[width=0.24\textwidth]{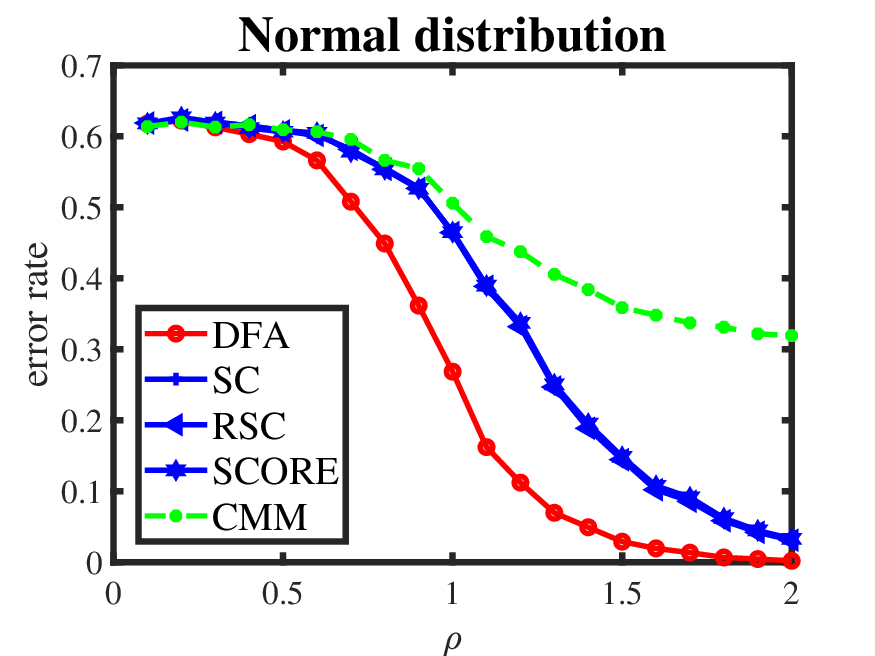}}
\subfigure[Changing $\sigma^{2}_{A}$]{\includegraphics[width=0.24\textwidth]{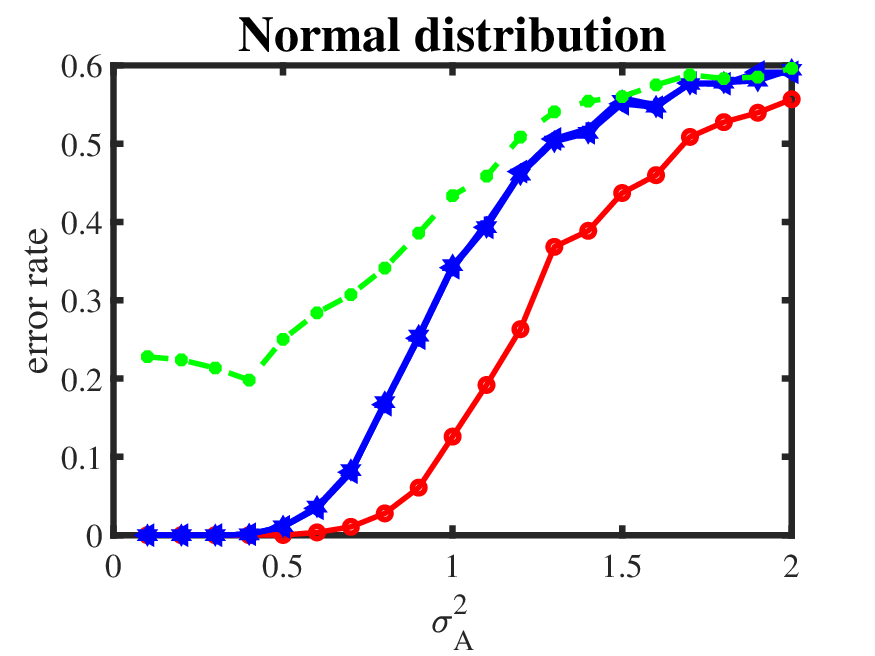}}
\subfigure[Changing $\sigma^{2}_{W}$]{\includegraphics[width=0.24\textwidth]{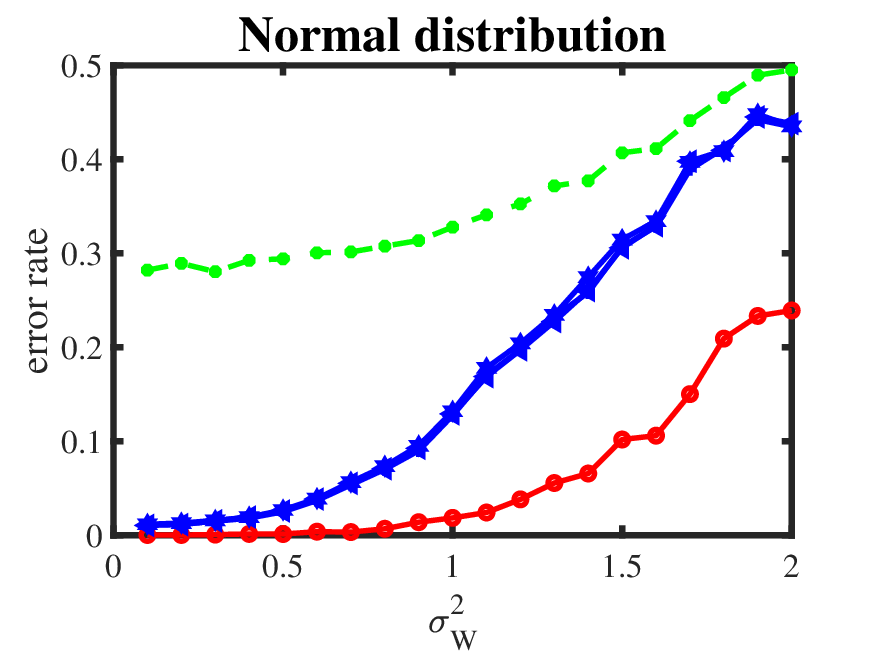}}
\subfigure[Changing $p$]{\includegraphics[width=0.24\textwidth]{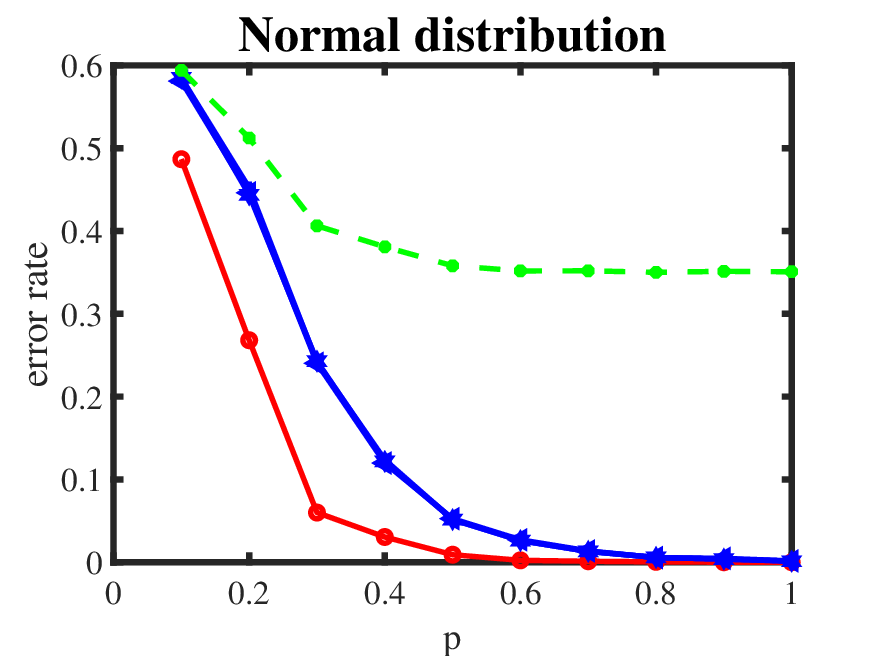}}
\subfigure[Changing $\rho$]{\includegraphics[width=0.24\textwidth]{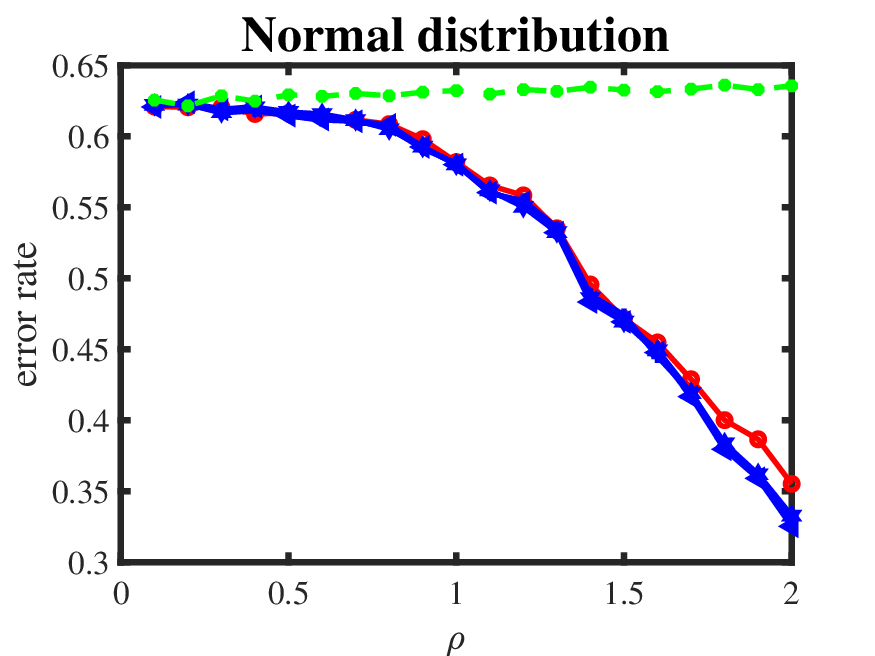}}
\subfigure[Changing $\sigma^{2}_{A}$]{\includegraphics[width=0.24\textwidth]{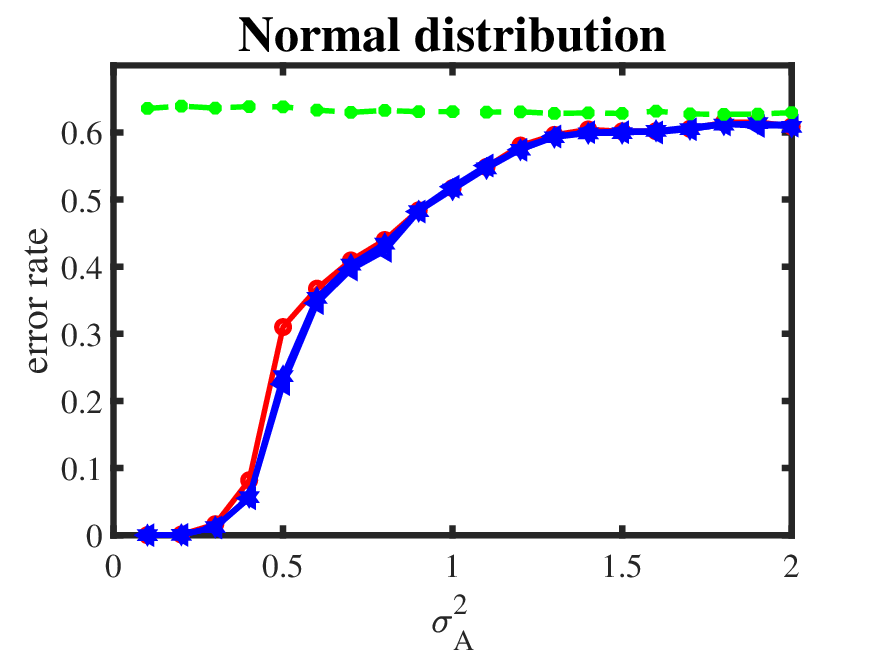}}
\subfigure[Changing $\sigma^{2}_{W}$]{\includegraphics[width=0.24\textwidth]{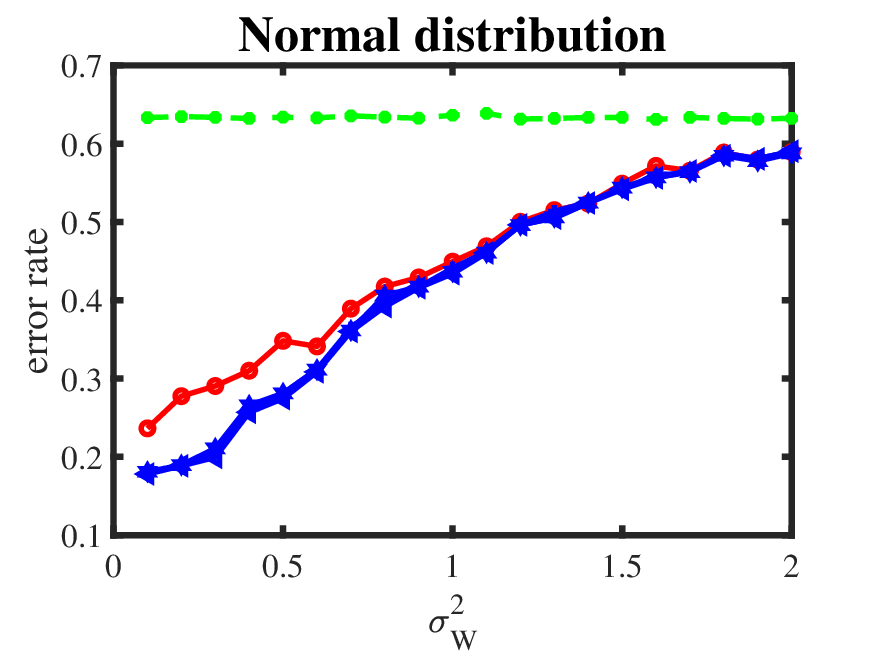}}
\subfigure[Changing $p$]{\includegraphics[width=0.24\textwidth]{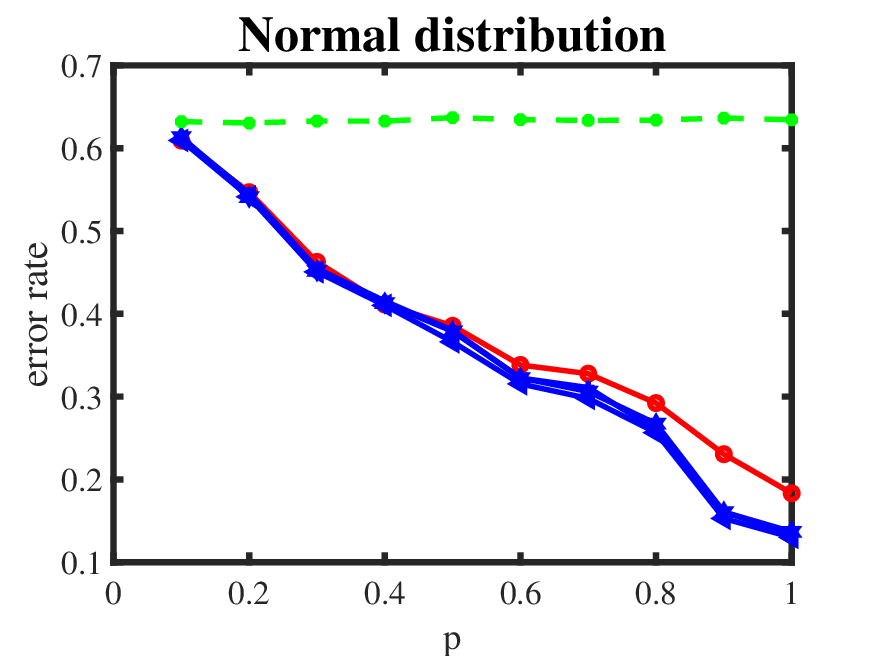}}
\caption{Numerical results of Experiment 2.}
\label{EX2} 
\end{figure}
When $A(i,j)\sim\mathrm{Normal}(\Omega(i,j),\sigma^{2}_{A})$ for some $\sigma^{2}_{A}>0$ for $i,j\in[n]$, by Example \ref{Normal}, we set $P$ as $p_{2}$ in Experiments 2[a]-2[d], and set it as $P_{3}$ in Experiments 2[e]-2[h].

\texttt{Experiment 2[a]: Changing $\rho$.} Let $\sigma^{2}_{A}=3, W$ be a zero matrix, and $\rho$ range in $\{0.1,0.2,\ldots,2\}$. Panel (a) of Figure \ref{EX2} shows the result. We see that, when $\rho$ increases, DFA performs better, and this is consistent with our findings in Example \ref{Normal}. Meanwhile, DFA performs best and CMM almost always fails to work in Experiment 2[a].

\texttt{Experiment 2[b]: Changing $\sigma^{2}_{A}$.} Let $\rho=0.4, W$ be a zero matrix, and $\sigma^{2}_{A}$ range in $\{0.1,0.2,\ldots,2\}$. Panel (b) of Figure \ref{EX2} shows the result, from which we see that decreasing $\sigma^{2}_{A}$ decreases error rates, and this is consistent with the findings in Example \ref{Normal} since a smaller $\sigma^{2}_{A}$ indicates a case that $A$ and $\Omega$ are closer. Meanwhile, DFA outperforms the other methods and CMM performs poorest for this sub-experiment.

\texttt{Experiment 2[c]: Changing $\sigma^{2}_{W}$.} Set $\rho=0.8$, $\sigma^{2}_{A}=1$, and let $\sigma^{2}_{W}$ range in $\{0.1, 0.2, \ldots, 2\}$. The result is displayed in panel (c) of Figure \ref{EX2}. We see that DFA performs poorer as $\sigma^{2}_{W}$ increases, and this phenomenon is consistent with our theoretical findings. Meanwhile, DFA performs best while CMM performs poorest here.

\texttt{Experiment 2[d]: Changing $p$.} In this sub-experiment, we study the influence of $p$ on the performances of different methods. Let $\rho=0.8,\sigma^{2}_{A}=1,$ and $W$ be a zero matrix. Let $p$ range in $\{0.1, 0.2, \ldots, 1\}$, and the result is displayed by the panel (d) of Figure \ref{EX2}. We see that all methods perform better as $p$ increases, DFA performs best, and CMM performs poorest.

For Experiments 2[e], 2[f], 2[g], and 2[h], except $P$, all parameters are set the same as Experiments 2[a], 2[b], 2[c], and 2[d], respectively. Panels (e)-(h) of Figure \ref{EX2} show the results. For DFA, SC, RSC, and SCORE, their performances for Experiments 2[e]-2[h] are similar to that of Experiments 2[a]-2[d]. It is interesting to find that CMM fails to detect communities for Experiments 2[e]-2[h]. Therefore, when all entries of $P$ are negative, spectral methods (DFA, SC, RSC, and SCORE) still work while the modularity maximization method CMM fails, and such a phenomenon is also found in Experiments 6 and 7.
\subsubsection{Experiment 3: Binomial distribution}
\begin{figure}
\centering
\subfigure[Changing $\rho$]{\includegraphics[width=0.24\textwidth]{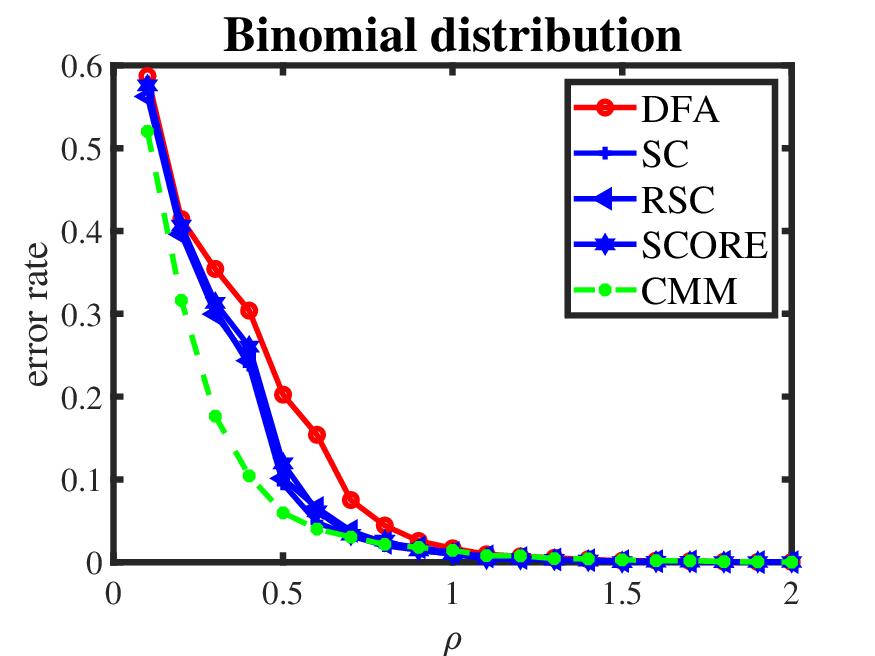}}
\subfigure[Changing $m$]{\includegraphics[width=0.24\textwidth]{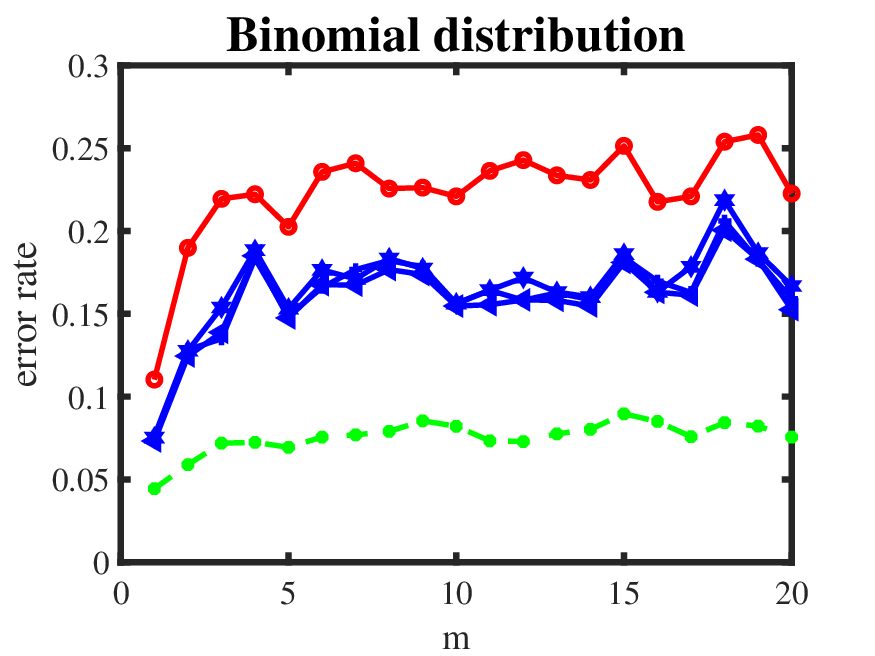}}
\subfigure[Changing $\sigma^{2}_{W}$]{\includegraphics[width=0.24\textwidth]{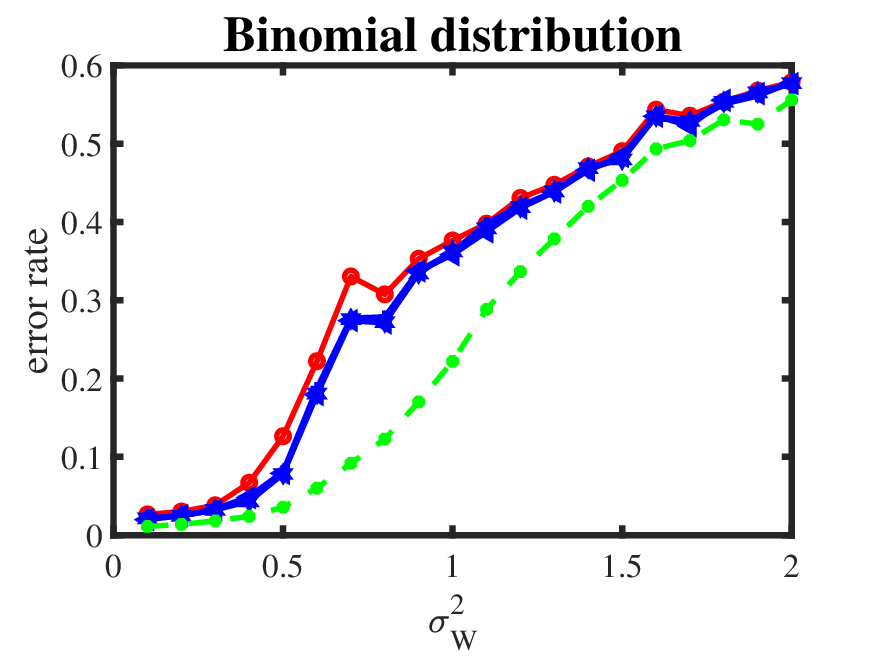}}
\subfigure[Changing $p$]{\includegraphics[width=0.24\textwidth]{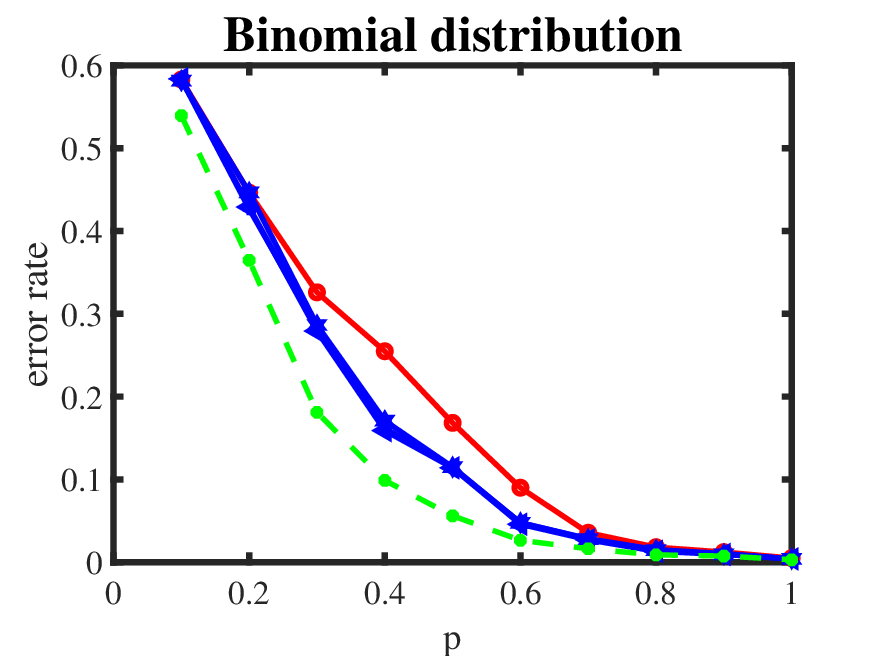}}
\caption{Numerical results of Experiment 3.}
\label{EX3} 
\end{figure}
When $A(i,j)\sim\mathrm{Binomial}(m,\frac{\Omega(i,j)}{m})$ for some positive integer $m$ for $i,j\in[n]$, by Example \ref{Binomial}, we set $P$ as $P_{1}$.

\texttt{Experiment 3[a]: Changing $\rho$.} Let $m=3$, and $\rho$ range in $\{0.1,0.2,\ldots,2\}$. Panel (a) of Figure \ref{EX3} shows the result, from which we see that all methods perform better as $\rho$ increases and this is consistent with our theoretical findings in Example \ref{Binomial}.

\texttt{Experiment 3[b]: Changing $m$.} Let $\rho=0.4$, and $m$ range in $\{1,2,\ldots,20\}$. Panel (b) of Figure \ref{EX3} shows the result. We see that all methods perform poorer as $m$ becomes larger. This can be explained as below: since $A$ follows Binomial distribution and $m$ is the number of trials in this experiment, a larger $m$ indicates increasing the variation of $A(i,j)$ for all nodes, hence it becomes harder to detect networks when $m$ is large. Meanwhile, CMM performs best and DFA performs poorest among these methods for this sub-experiment.

\texttt{Experiment 3[c]: Changing $\sigma^{2}_{W}$.} Set $\rho=0.8$, $m=3$, and let $\sigma^{2}_{W}$ range in $\{0.1, 0.2, \ldots, 2\}$. The result is shown in the third panel of Figure \ref{EX3}, and the conclusion is similar to that of Experiment 1[b].

\texttt{Experiment 3[d]: Changing $p$.} Let $\rho=0.8, m=3$, and $W$ be a zero matrix. Let $p$ range in $\{0.1, 0.2, \ldots, 1\}$, and the result is displayed by the last panel of Figure \ref{EX3}. The conclusion is similar to that of Experiment 1[c].
\subsubsection{Experiment 4: Poisson distribution}
\begin{figure}
\centering
\subfigure[Changing $\rho$]{\includegraphics[width=0.325\textwidth]{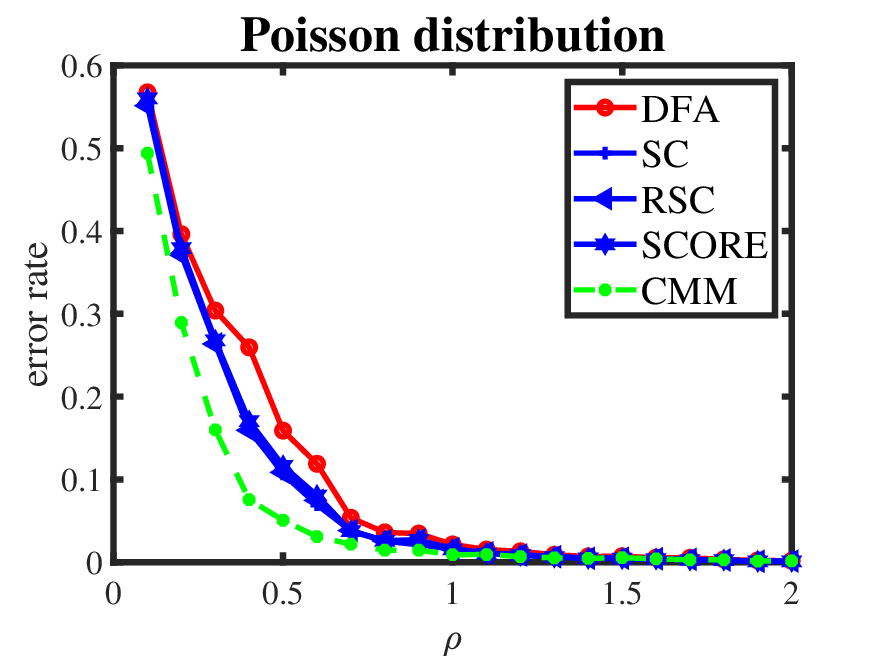}}
\subfigure[Changing $\sigma^{2}_{W}$]{\includegraphics[width=0.325\textwidth]{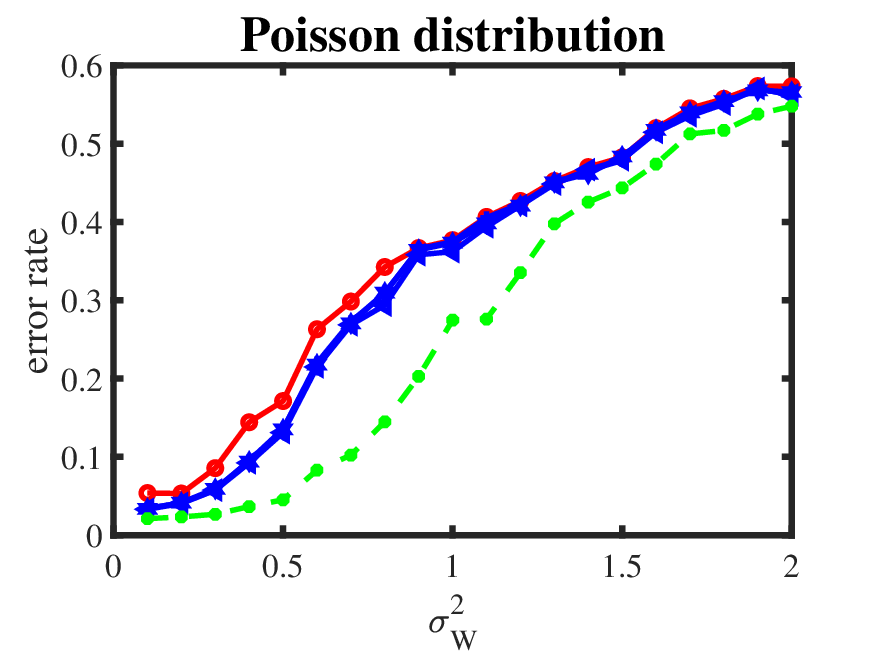}}
\subfigure[Changing $p$]{\includegraphics[width=0.325\textwidth]{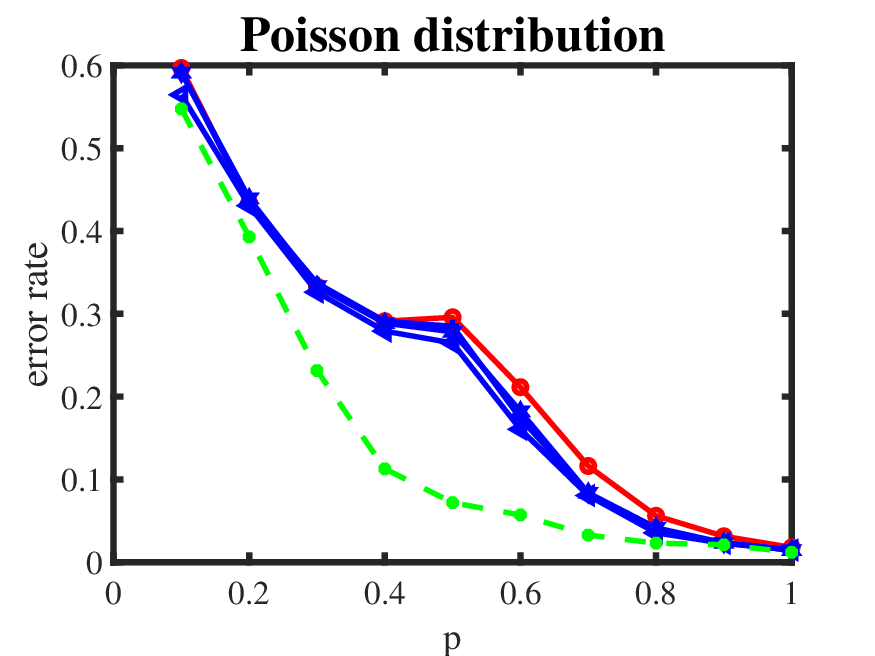}}
\caption{Numerical results of Experiment 4.}
\label{EX4} 
\end{figure}
When $A(i,j)\sim\mathrm{Poisson}(\Omega(i,j))$ for $i,j\in[n]$, by Example \ref{Poisson}, we set $P$ as $P_{1}$.

\texttt{Experiment 4[a]: Changing $\rho$.} Let $\rho$ range in $\{0.1,0.2,\ldots,2\}$. Panel (a) of Figure \ref{EX4} shows the result, and the conclusion is similar to that of Experiments 1[a], 2[a], and 3[a].

\texttt{Experiment 4[b]: Changing $\sigma^{2}_{W}$.} Set $\rho=0.8$ and let $\sigma^{2}_{W}$ range in $\{0.1, 0.2, \ldots, 2\}$. Panel (b) of Figure \ref{EX4} shows the results, with a conclusion similar to that of Experiments 1[b], 2[c], and 3[c].

\texttt{Experiment 4[c]: Changing $p$.} Let $\rho=0.8$ and $W$ be a zero matrix. Let $p$ range in $\{0.1, 0.2, \ldots, 1\}$, and the result is displayed by the panel (c) of Figure \ref{EX4}. The conclusion is similar to that of Experiment 1[c].
\subsubsection{Experiment 5: Exponential distribution}
When $A(i,j)\sim \mathrm{Exponential}(\frac{1}{\Omega(i,j)})$ for $i,j\in[n]$, by Example \ref{Exponential}, we set $P$ as $P_{1}$.
\begin{figure}
\centering
\subfigure[Changing $\rho$]{\includegraphics[width=0.325\textwidth]{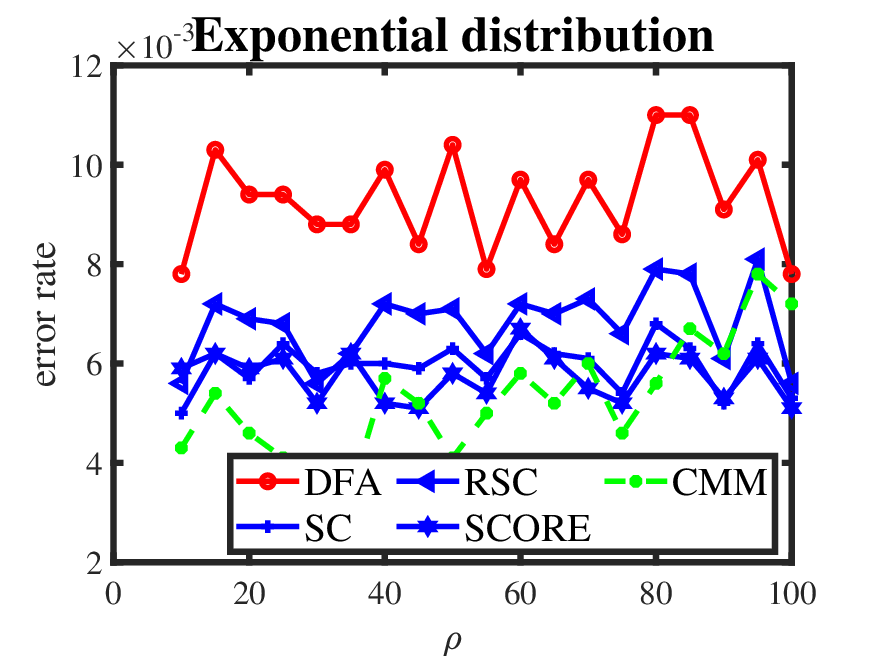}}
\subfigure[Changing $\sigma^{2}_{W}$]{\includegraphics[width=0.325\textwidth]{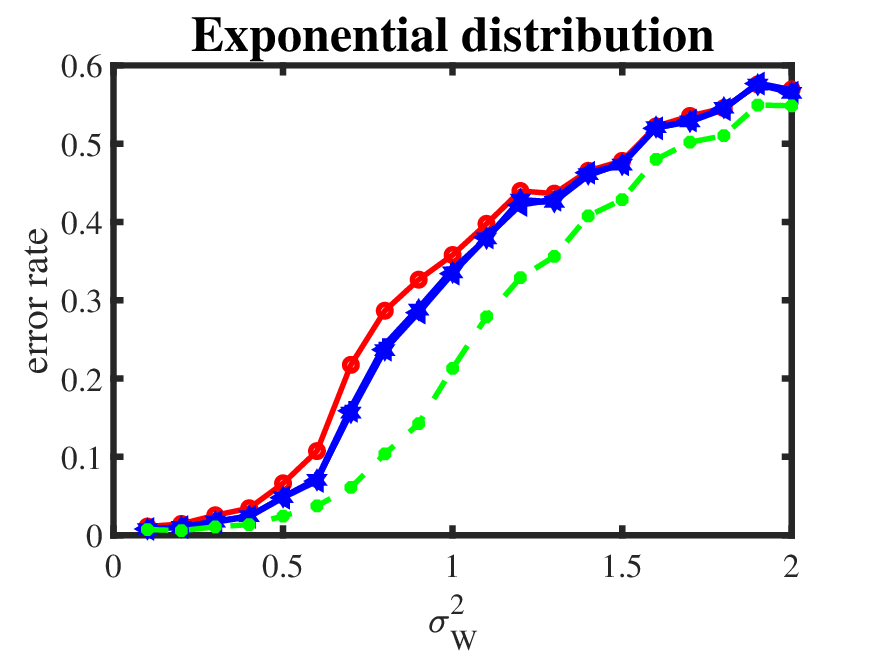}}
\subfigure[Changing $p$]{\includegraphics[width=0.325\textwidth]{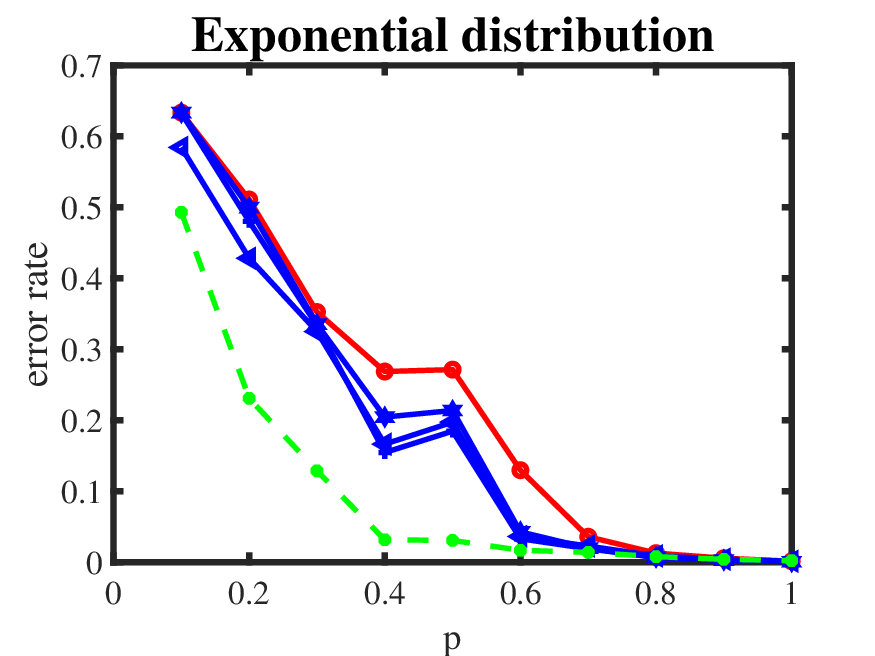}}
\caption{Numerical results of Experiment 5.}
\label{EX5} 
\end{figure}

\texttt{Experiment 5[a]: Changing $\rho$.} Let $\rho$ range in $\{10,15,\ldots,100\}$. Panel (a) of Figure \ref{EX5} shows the result. We see that increasing $\rho$ has no significant influence on the performances of all methods, and this supports our theoretical findings in Example \ref{Exponential}. Meanwhile, all methods enjoy satisfactory performances for their small error rates in this experiment.

\texttt{Experiment 5[b]: Changing $\sigma^{2}_{W}$.} Set $\rho=0.8$ and let $\sigma^{2}_{W}$ range in $\{0.1, 0.2, \ldots, 2\}$. Panel (b) of Figure \ref{EX5} shows that the result is similar to previous experiments.

\texttt{Experiment 5[c]: Changing $p$.} Let $\rho=1$ and $W$ be a zero matrix. Let $p$ range in $\{0.1, 0.2, \ldots, 1\}$, and the result is displayed by the panel (c) of Figure \ref{EX5}. The conclusion is similar to previous experiments.
\subsubsection{Experiment 6: Uniform distribution}
For Uniform distribution, we set $K=3, n=100$. For Experiments 6[a], 6[b], and 6[c], we consider Uniform-Case (1) when $A(i,j)\sim\mathrm{Uniform}[0,2\Omega(i,j)]$. For Uniform-Case (1), by Example \ref{Uniform}, we set $P$ as $P_{1}$.
\begin{figure}
\centering
\subfigure[Changing $\rho$]{\includegraphics[width=0.325\textwidth]{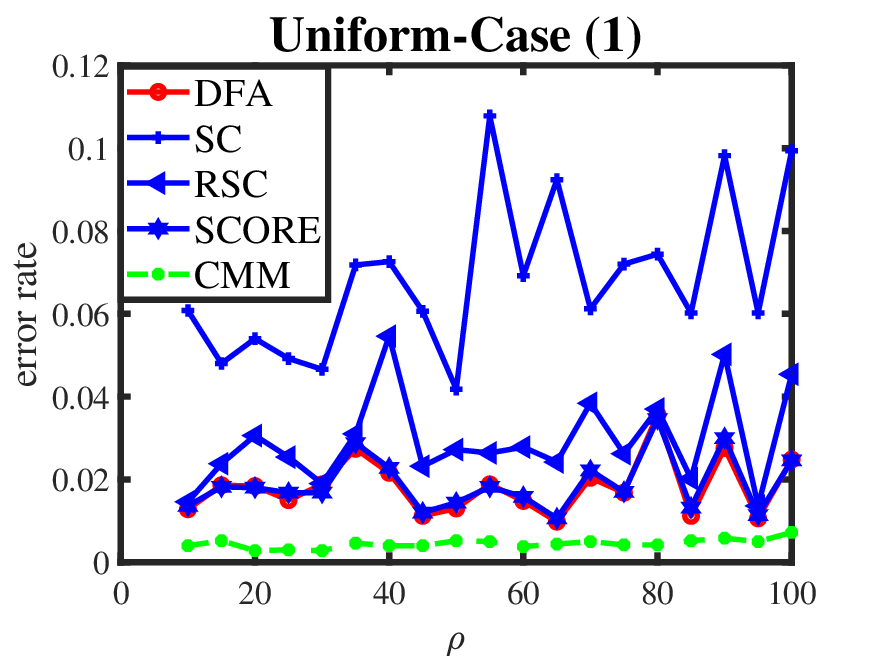}}
\subfigure[Changing $\sigma^{2}_{W}$]{\includegraphics[width=0.325\textwidth]{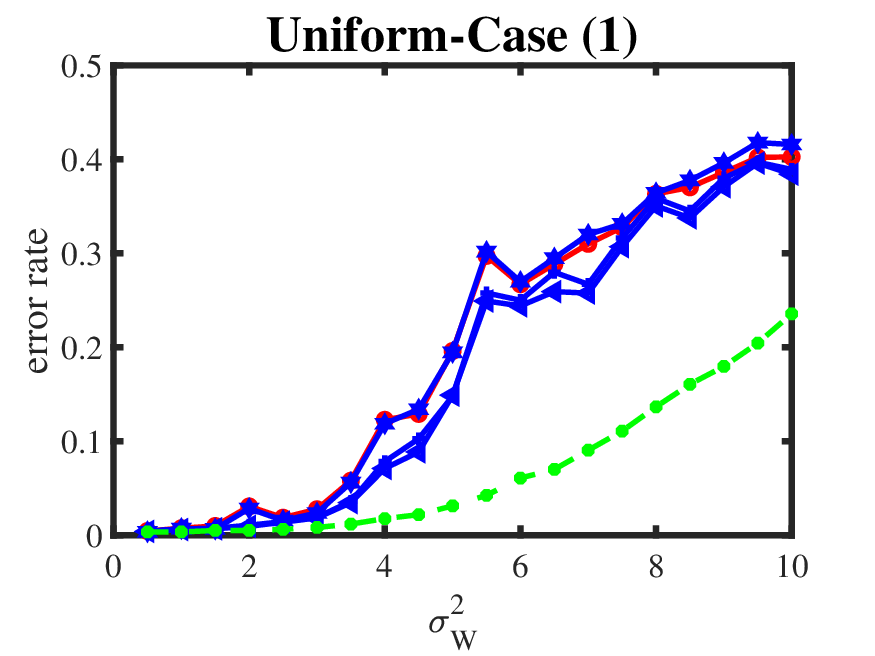}}
\subfigure[Changing $p$]{\includegraphics[width=0.325\textwidth]{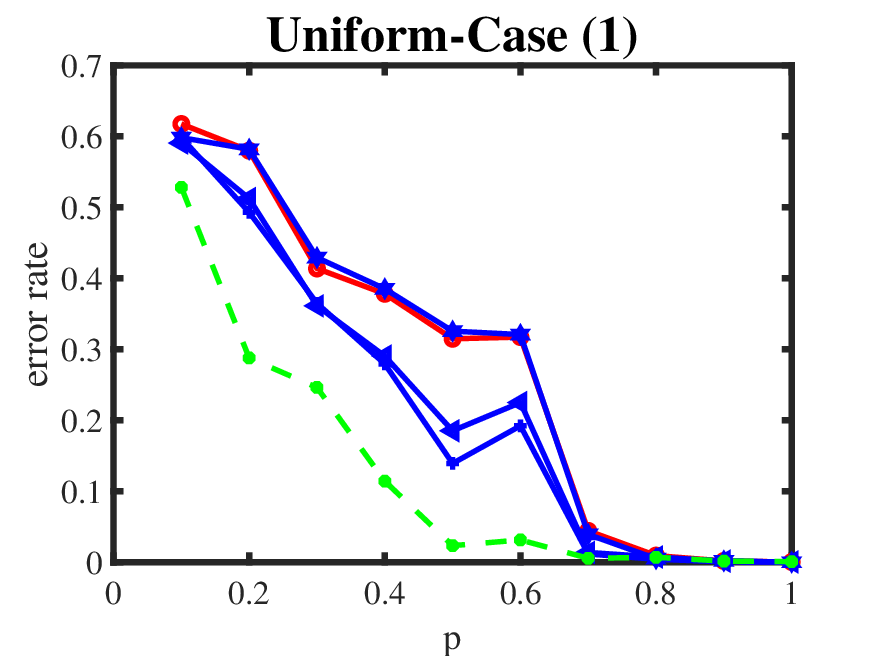}}
\subfigure[Changing $\rho$]{\includegraphics[width=0.325\textwidth]{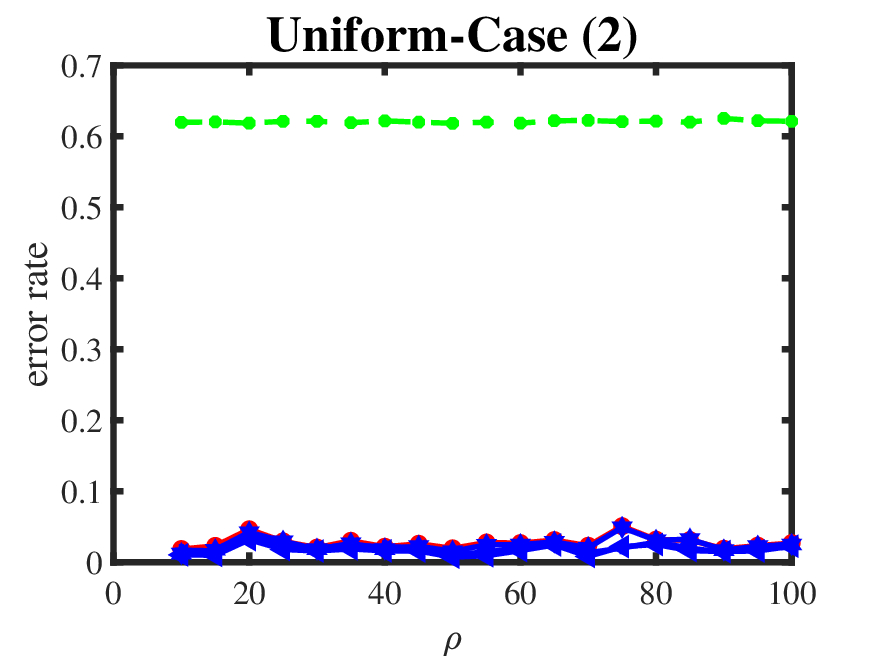}}
\subfigure[Changing $\sigma^{2}_{W}$]{\includegraphics[width=0.325\textwidth]{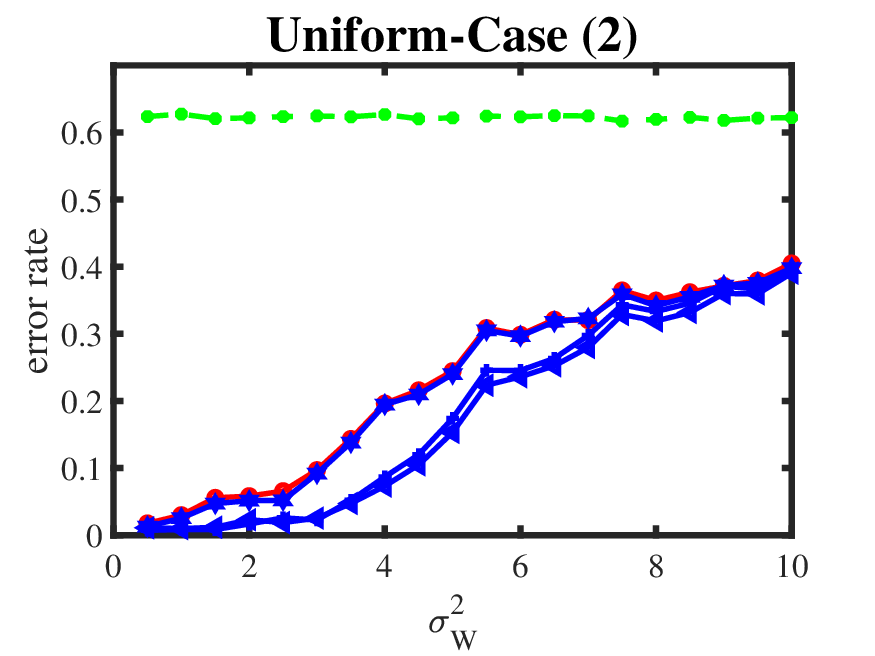}}
\subfigure[Changing $p$]{\includegraphics[width=0.325\textwidth]{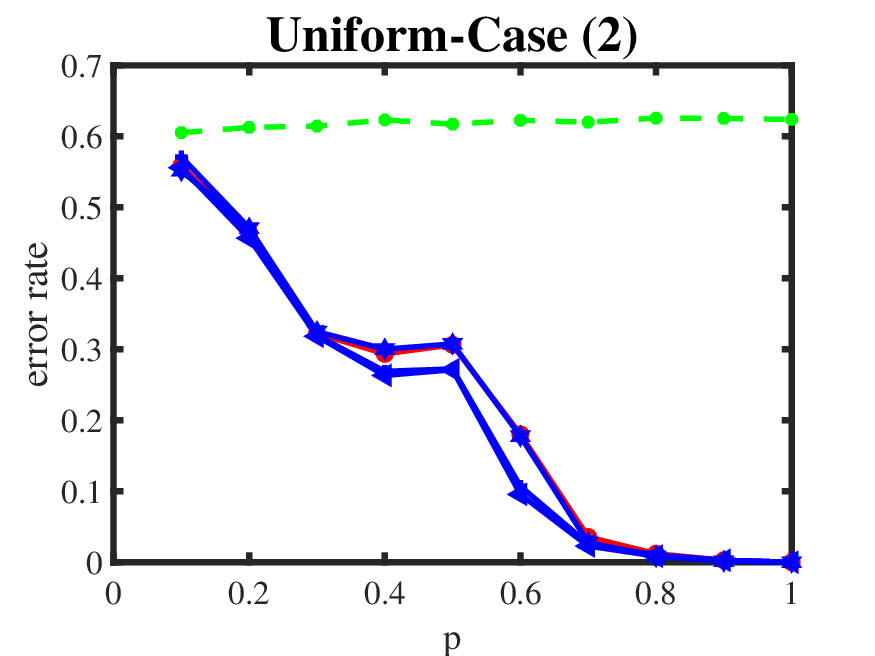}}
\caption{Numerical results of Experiment 6. }
\label{EX6} 
\end{figure}

\texttt{Experiment 6[a]: Changing $\rho$.} Let $\rho$ range in $\{10,15,\ldots,100\}$. Panel (a) of Figure \ref{EX6} shows the result. We see that increasing $\rho$ has no significant influence on the performances of these methods and this is consistent with our theoretical analysis in Example \ref{Uniform}. Meanwhile, all methods perform satisfactorily here for their small error rates.

\texttt{Experiment 6[b]: Changing $\sigma^{2}_{W}$.} Set $\rho=10$ and let $\sigma^{2}_{W}$ range in $\{0.5, 1, \ldots, 10\}$. Panel (b) of Figure \ref{EX6} displays the result. The conclusion is similar to that of Experiment 1[b].

\texttt{Experiment 6[c]: Changing $p$.} Let $\rho=10$ and $W$ be a zero matrix. Let $p$ range in $\{0.1, 0.2, \ldots, 1\}$. The result can be found in panel (c) of Figure \ref{EX6} and it is similar to previous experiments.
\begin{figure}
\centering
\subfigure[Changing $\rho$]{\includegraphics[width=0.325\textwidth]{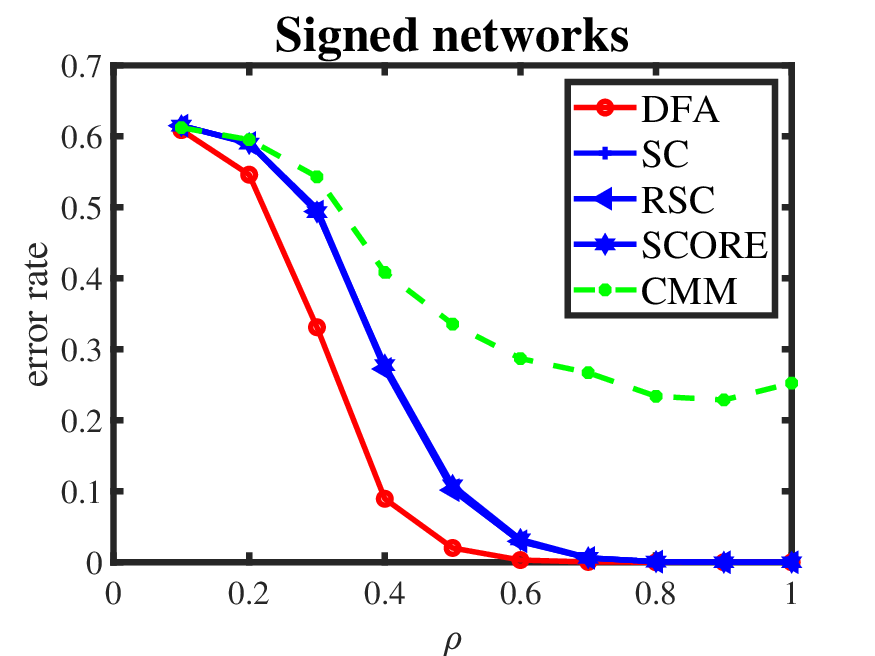}}
\subfigure[Changing $\sigma^{2}_{W}$]{\includegraphics[width=0.325\textwidth]{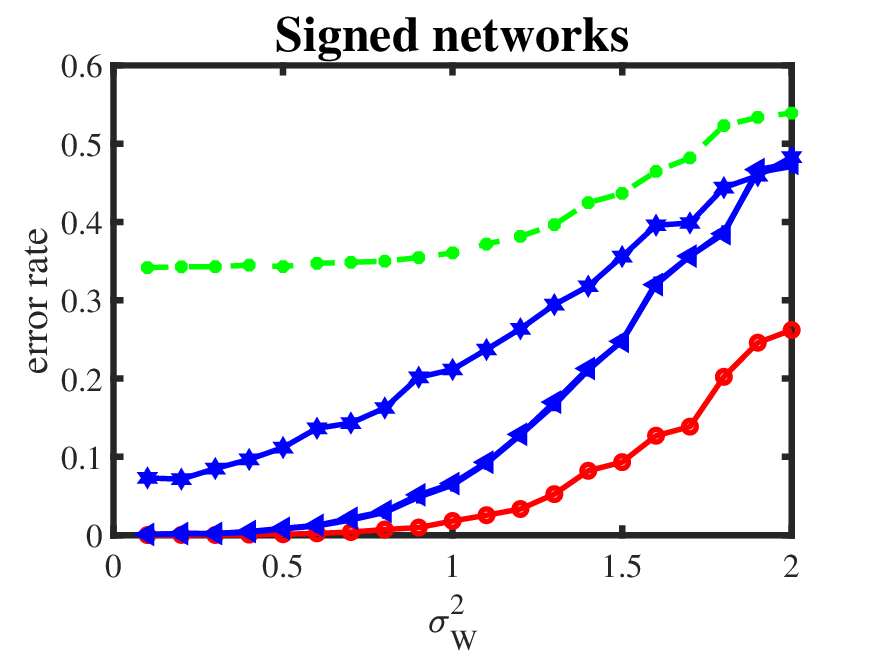}}
\subfigure[Changing $p$]{\includegraphics[width=0.325\textwidth]{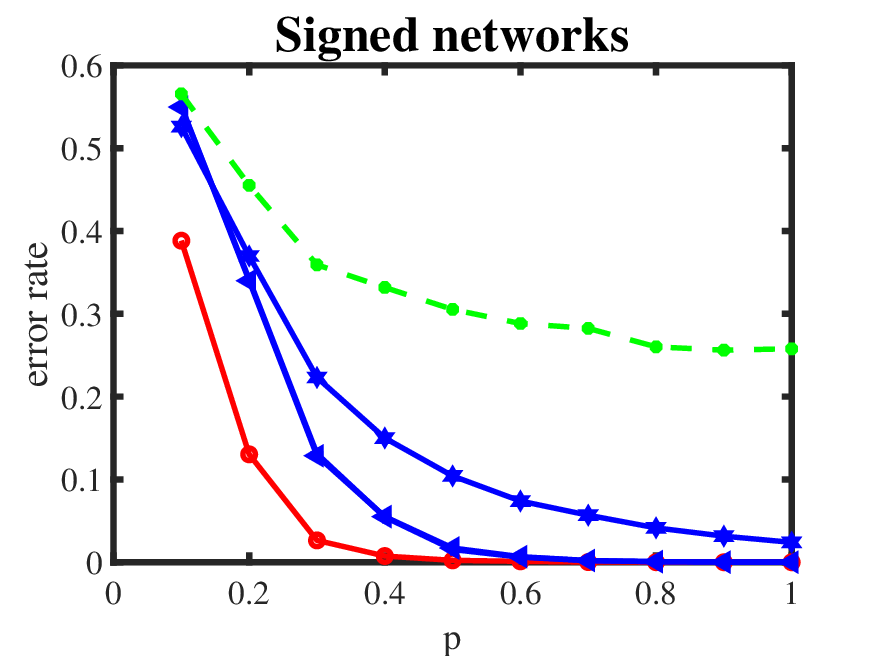}}
\subfigure[Changing $\rho$]{\includegraphics[width=0.325\textwidth]{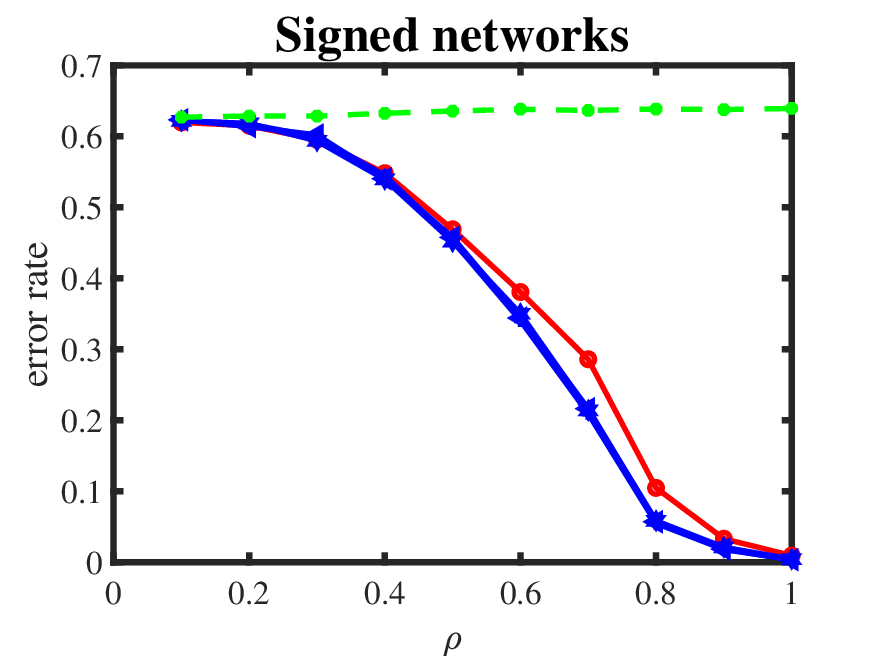}}
\subfigure[Changing $\sigma^{2}_{W}$]{\includegraphics[width=0.325\textwidth]{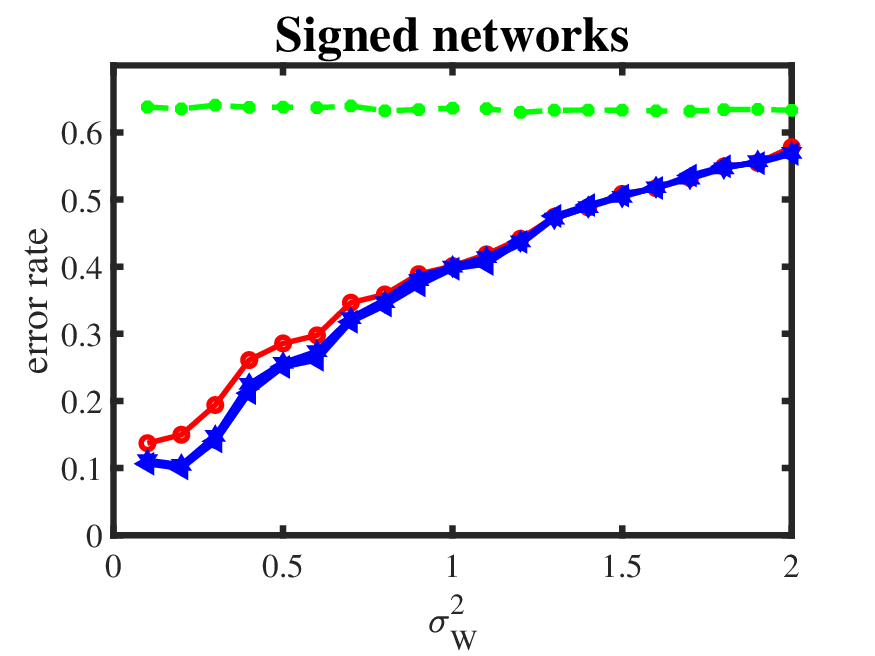}}
\subfigure[Changing $p$]{\includegraphics[width=0.325\textwidth]{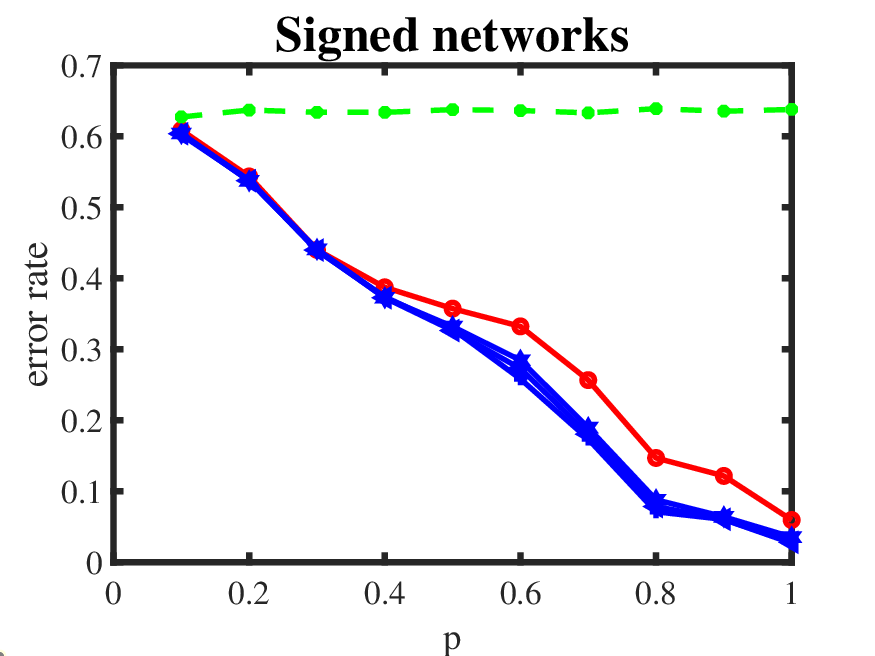}}
\caption{Numerical results of Experiment 7. }
\label{EX7} 
\end{figure}

For Experiments 6[d], 6[e], and 6[f], we consider Uniform-Case (2) when $A(i,j)\sim\mathrm{Uniform}[2\Omega(i,j),0]$. For Uniform-Case (2), by Example \ref{Uniform}, we set $P$ as $P_{3}$.

\texttt{Experiment 6[d]: Changing $\rho$.} Let $\rho$ range in $\{10,15,\ldots,100\}$.

\texttt{Experiment 6[e]: Changing $\sigma^{2}_{W}$.} Set $\rho=10$ and let $\sigma^{2}_{W}$ range in $\{0.5, 1, \ldots, 10\}$.

\texttt{Experiment 6[f]: Changing $p$.} Let $\rho=10$ and $W$ be a zero matrix. Let $p$ range in $\{0.1, 0.2, \ldots, 1\}$.

Panels (d), (e), and (f) of Figure \ref{EX6} show the results of Experiments 6[d], 6[e], and 6[f], respectively. The conclusions are similar to that of Experiments 2[e], 2[g], and 2[h], where spectral methods (DFA, SC, RSC, and SCORE) still work while the modularity maximization method CMM fails when all entries of $P$ are negative for these experiments.
\subsubsection{Experiment 7: Signed networks}
For signed networks when $\mathbb{P}(A(i,j)=1)=\frac{1+\Omega(i,j)}{2}$ and $\mathbb{P}(A(i,j)=-1)=\frac{1-\Omega(i,j)}{2}$ for $i,j\in[n]$, by Example \ref{Signed}, we set $P$ as $p_{2}$ in Experiments 7[a]-7[c], and set it as $P_{3}$ in Experiments 7[d]-7[f].

\texttt{Experiment 7[a]: Changing $\rho$.} Let $\rho$ range in $\{0.1,0.2,\ldots,1\}$. Panel (a) of Figure \ref{EX7} shows the result. We see that increasing $\rho$ decreases the error rate of DFA for signed networks and this is consistent with our findings in Example \ref{Signed}. Meanwhile, DFA outperforms the other four methods and CMM performs poorest for this experiment.

\texttt{Experiment 7[b]: Changing $\sigma^{2}_{W}$.} Set $\rho=0.8$ and let $\sigma^{2}_{W}$ range in $\{0.1, 0.2, \ldots, 2\}$. Panel (b) of Figure \ref{EX7} shows the result which is similar to Experiment 2[c].

\texttt{Experiment 7[c]: Changing $p$.} Let $\rho=0.8$ and $W$ be a zero matrix. Let $p$ range in $\{0.1, 0.2, \ldots, 1\}$, and the result is displayed by the panel (c) of Figure \ref{EX7}. We see that all methods enjoy better performances when there are lesser missing edges and DFA performs best here.

For Experiments 7[d], 7[e], and 7[f], except $P$, all parameters are set the same as Experiments 7[a], 7[b], and 7[c], respectively. Panels (d), (e), and (f) of Figure \ref{EX7} show the results of Experiments 7[d], 7[e], and 7[f], respectively. We see that the performances of DFA, SC, RSC, and SCORE are similar to that of Experiments 7[a], 7[b], and 7[c] while CMM fails to detect communities for Experiments 7[d], 7[e], and 7[f].
\subsubsection{Experiment 8: Changing $\sigma_{K}(P)$.}
\begin{figure}
\centering
\subfigure[]{\includegraphics[width=0.325\textwidth]{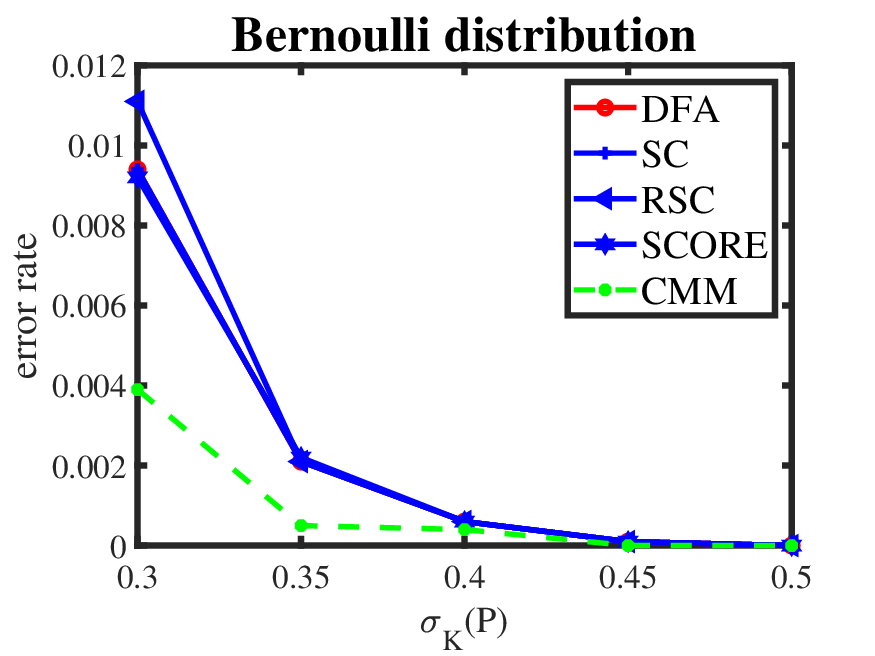}}
\subfigure[]{\includegraphics[width=0.325\textwidth]{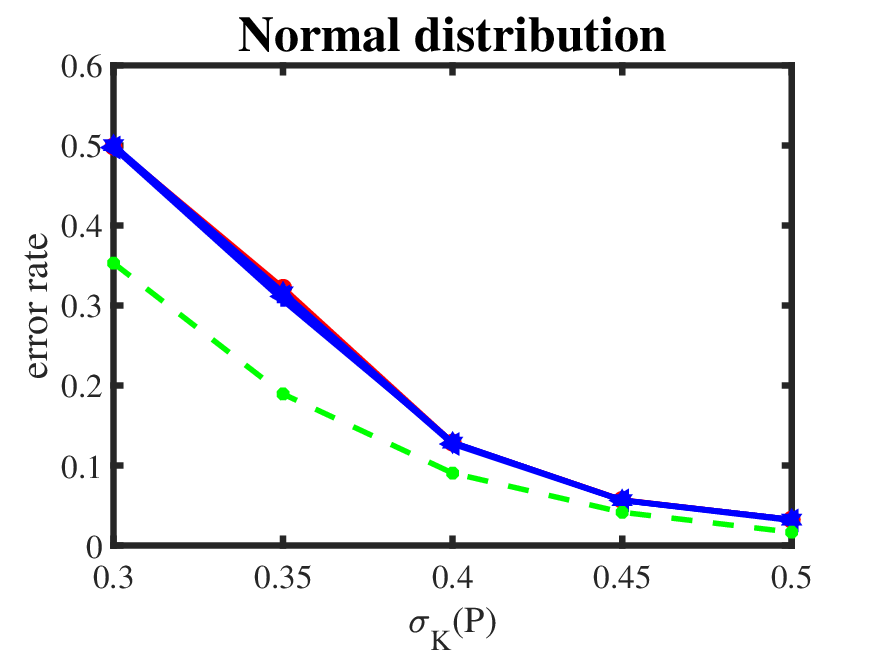}}
\subfigure[]{\includegraphics[width=0.325\textwidth]{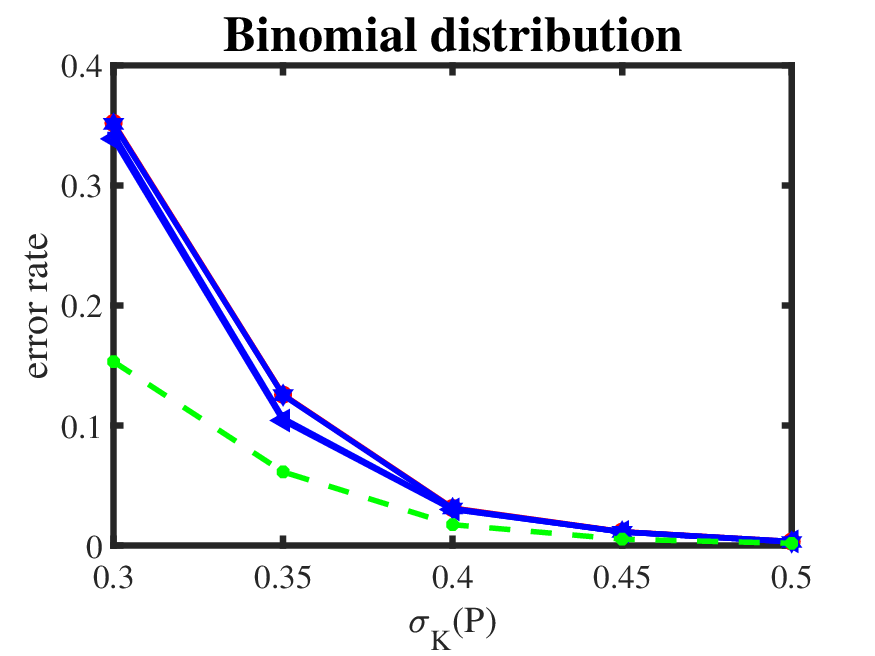}}
\subfigure[]{\includegraphics[width=0.325\textwidth]{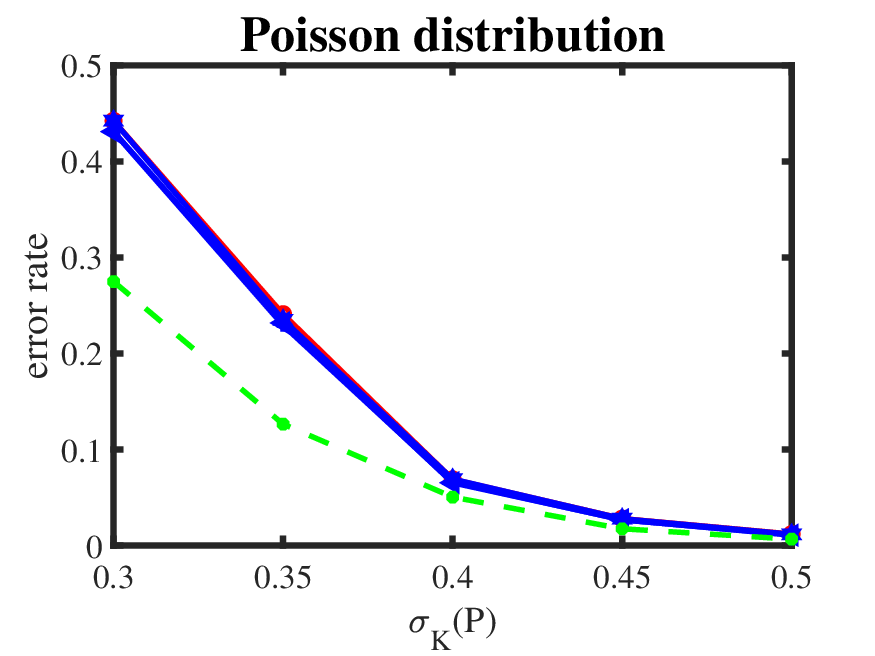}}
\subfigure[]{\includegraphics[width=0.325\textwidth]{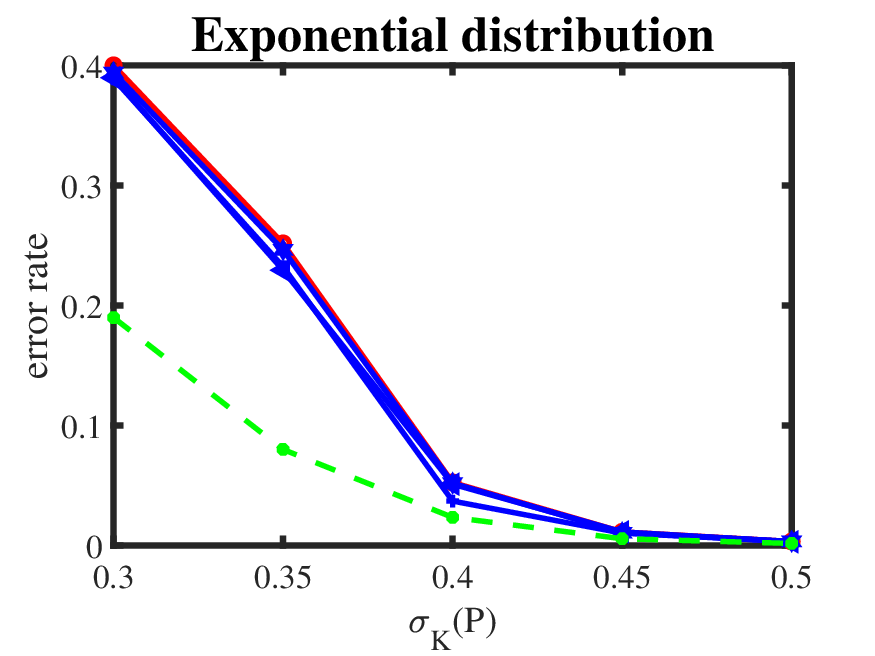}}
\subfigure[]{\includegraphics[width=0.325\textwidth]{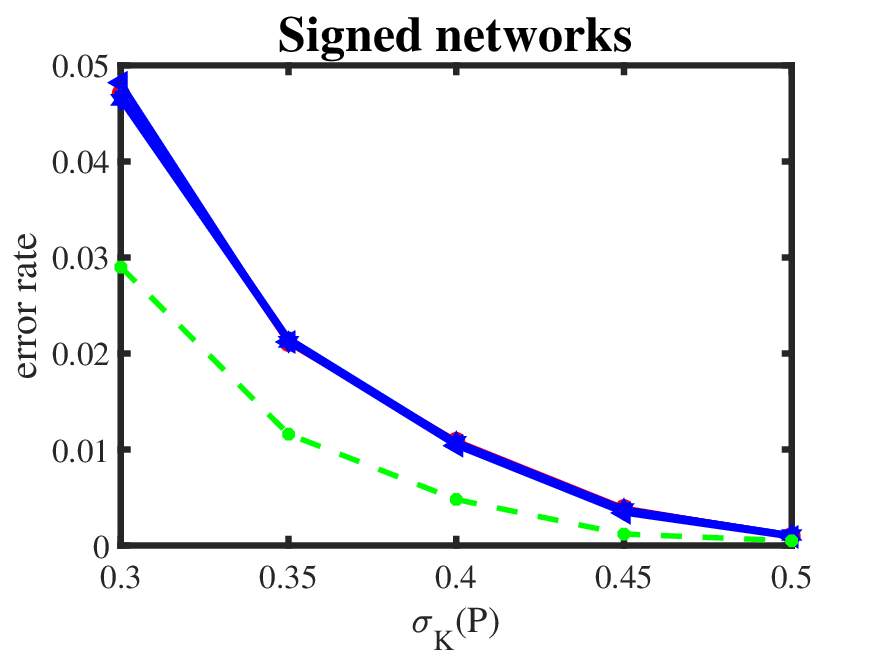}}
\caption{Numerical results of Experiment 8.}
\label{EX8} 
\end{figure}

\begin{figure*}
\centering
\subfigure[$\hat{A}_{\mathcal{A}}$ of Set-up 1]{\includegraphics[width=0.4\textwidth]{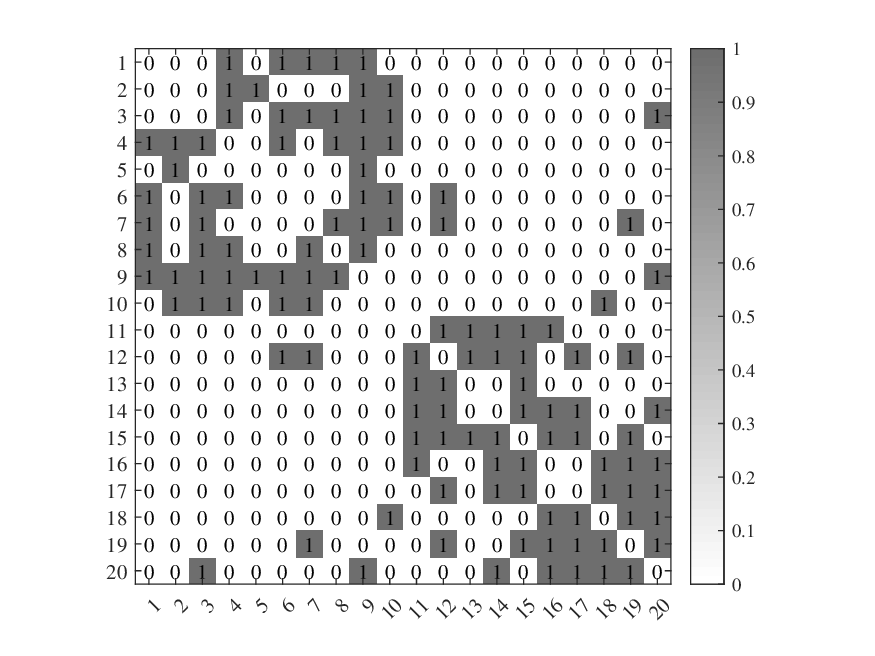}}
\subfigure[$\hat{A}_{\mathcal{A}}$ of Set-up 2]{\includegraphics[width=0.4\textwidth]{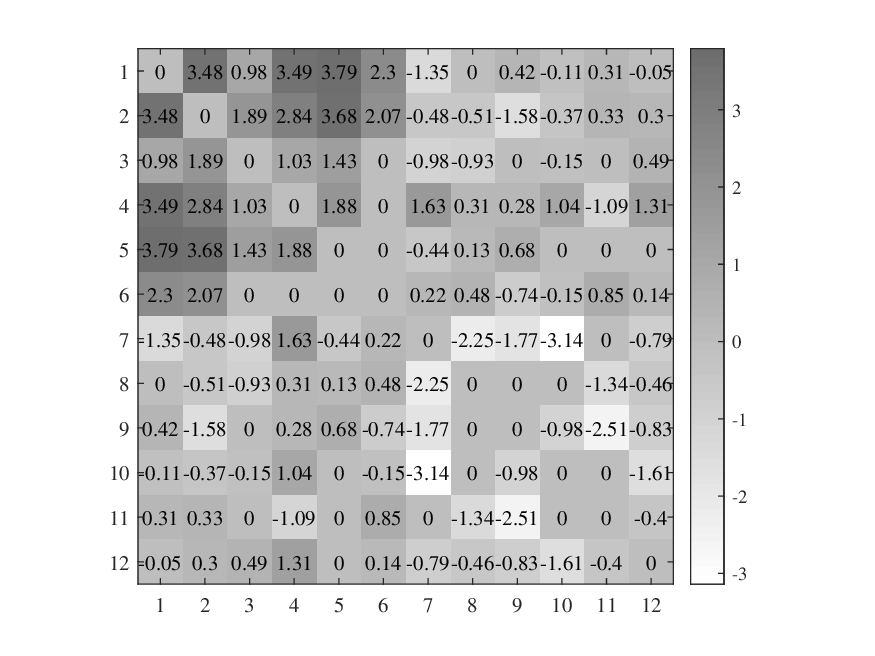}}
\subfigure[$\hat{A}_{\mathcal{A}}$ of Set-up 3]{\includegraphics[width=0.4\textwidth]{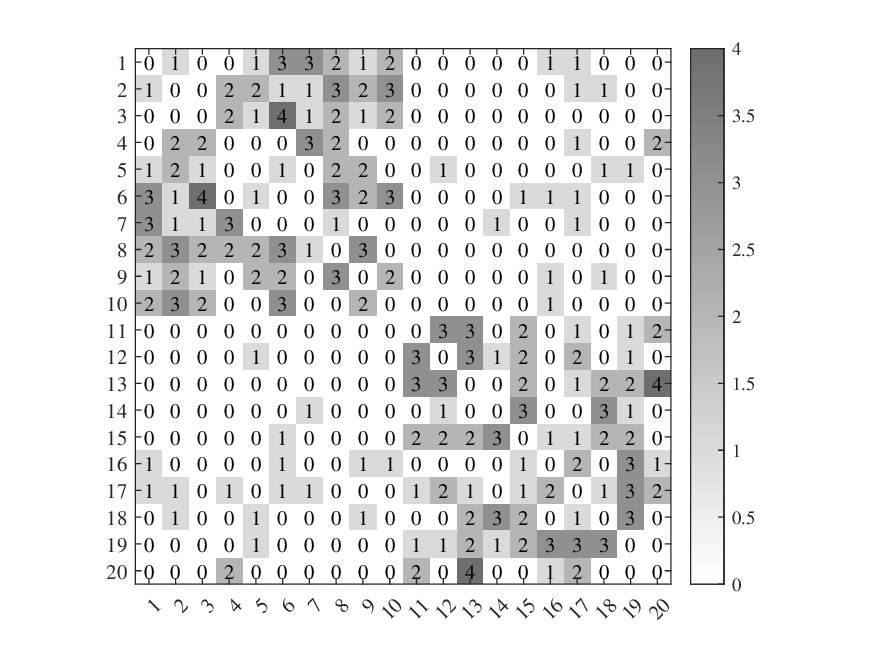}}
\subfigure[$\hat{A}_{\mathcal{A}}$ of Set-up 4]{\includegraphics[width=0.4\textwidth]{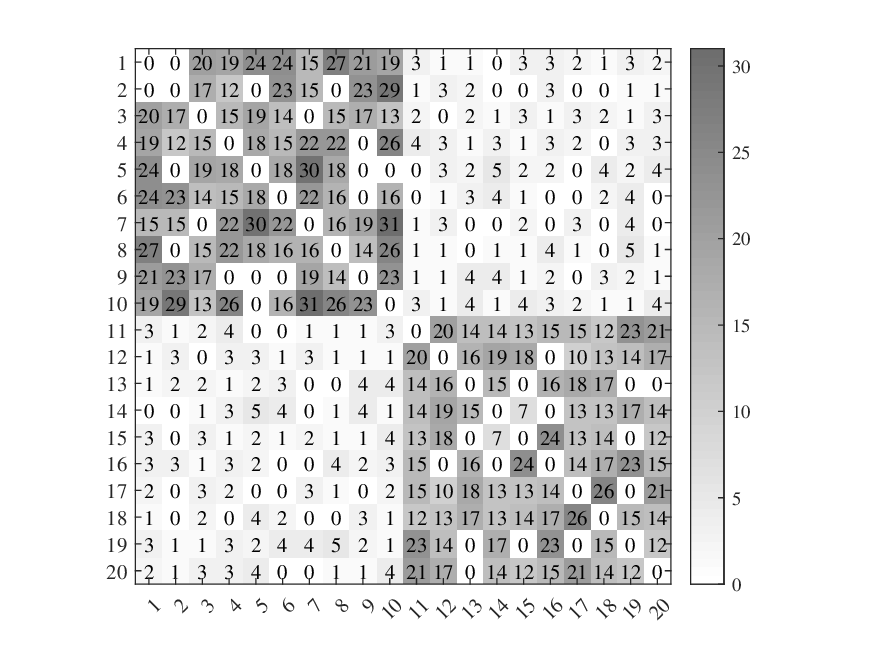}}
\subfigure[$\hat{A}_{\mathcal{A}}$ of Set-up 5]{\includegraphics[width=0.4\textwidth]{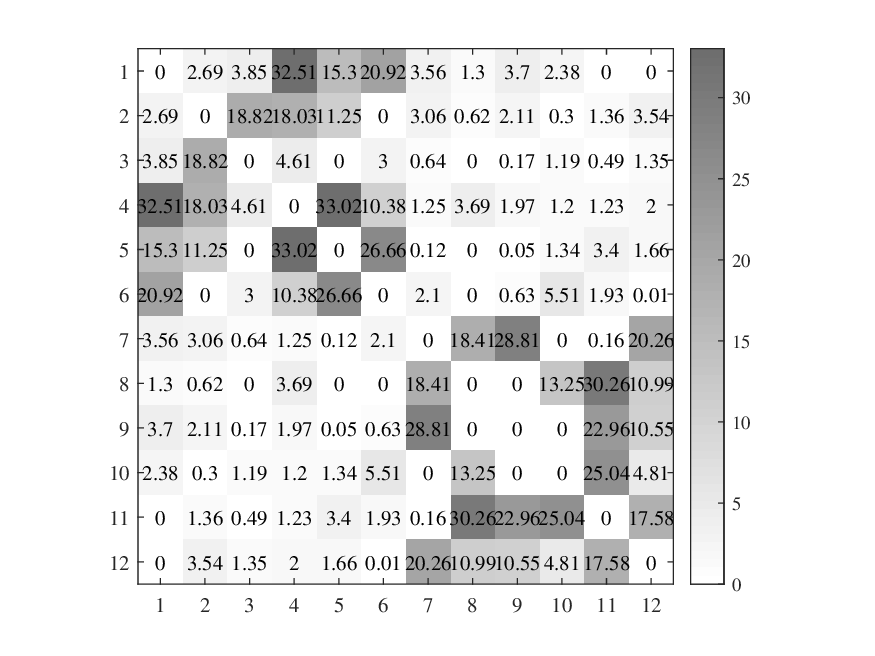}}
\subfigure[$\hat{A}_{\mathcal{A}}$ of Set-up 6]{\includegraphics[width=0.4\textwidth]{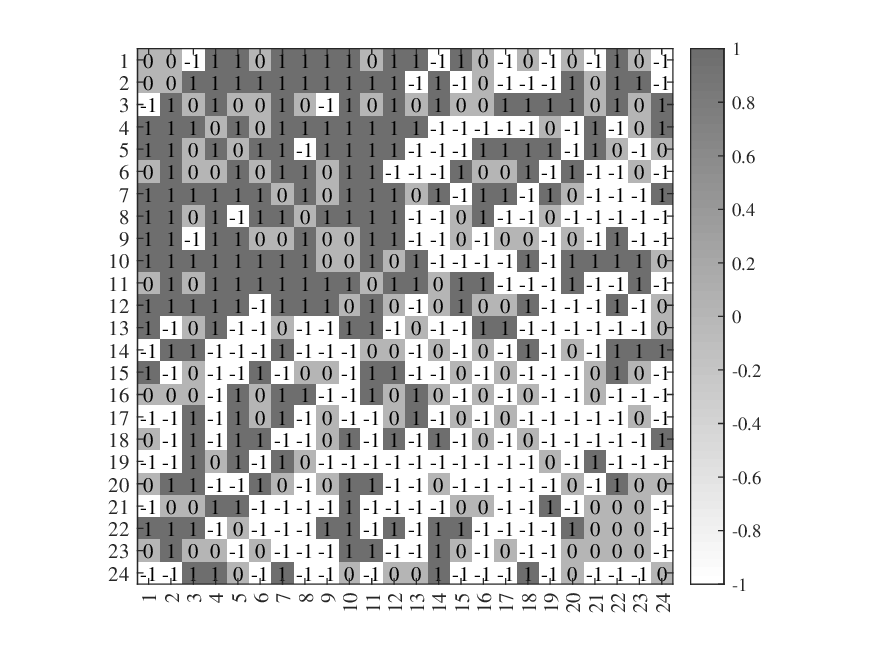}}
\caption{Illustration for weighted networks' adjacency matrices $\hat{A}_{\mathcal{A}}$ generated by (DFM+$G(n,p)$) when $p=0.8$. In panels (b) and (e), we keep all elements of $\hat{A}_{\mathcal{A}}$ in two decimals for visualization. Since $p=0.8<1$, some non-diagonal entries of adjacency matrices in all panels are zero and this means that there are some missing edges for these adjacency matrices.}
\label{AF}
\end{figure*}
Set $n=200, \rho=1, p=0.8$, and $P$ as
 \[P=\begin{bmatrix}
 1&1-\beta&1-\beta\\
 1-\beta&1&1-\beta\\
 1-\beta&1-\beta&1\\
 \end{bmatrix}.
 \]
Let $\beta\in\{0.3,0.35,0.4,0.45,0.5\}$. Since $\sigma_{K}(P)=\beta$, increasing $\beta$ should decrease DFA's error rate by Theorem \ref{MainDFA}, and we aim to verify this result in this experiment. Let $W$ be a zero matrix. This experiment has six sub-experiments.

\texttt{Experiment 8[a].} Let $A(i,j)\sim\mathrm{Bernoulli}(\Omega(i,j))$ for $i,j\in[n]$.

\texttt{Experiment 8[b].} Let $A(i,j)\sim\mathrm{Normal}(\Omega(i,j),1)$ for $i,j\in[n]$.

\texttt{Experiment 8[c].} Let $A(i,j)\sim\mathrm{Binomial}(3,\frac{\Omega(i,j)}{3})$ for $i,j\in[n]$.

\texttt{Experiment 8[d].} Let $A(i,j)\sim\mathrm{Poisson}(\Omega(i,j))$ for $i,j\in[n]$.

\texttt{Experiment 8[e].} Let $A(i,j)\sim\mathrm{Exponential}(\frac{1}{\Omega(i,j)})$ for $i,j\in[n]$.

\texttt{Experiment 8[f].} Let $\mathbb{P}(A(i,j)=1)=\frac{1+\Omega(i,j)}{2}$ and $\mathbb{P}(A(i,j)=-1)=\frac{1-\Omega(i,j)}{2}$ for $i,j\in[n]$.

Figure \ref{EX8} shows the results. We see that all methods perform better when $\beta$ becomes larger and this is consistent with our theoretical findings in Theorem \ref{MainDFA}.
\begin{Remark}
\begin{table}[h!]
\footnotesize
	\centering
	\caption{Error rates of methods used in this paper for adjacency matrices generated by Set-ups 1-6.}
	\label{TableSetUp}
	\begin{tabular}{cccccccccccc}
\hline\hline&Set-up 1&Set-up 2&Set-up 3&Set-up 4&Set-up 5&Set-up 6\\
\hline
DFA&0&0&0&0&0&0\\
SC&0&0.3333&0&0&0&0.3333\\
RSC&0&0.3333&0&0&0&0.2083\\
SCORE&0&0.3333&0&0&0&0.2083\\
CMM&0&0.0833&0&0&0&0.3333\\
\hline\hline
\end{tabular}
\end{table}
For visuality, we plot $\hat{A}_{\mathcal{A}}$ generated from (DFM+$G(n,p)$) (so we do not consider noise matrix here) in this remark. We let $K=2, p=0.8$, nodes in $\{1,2,\ldots, n/2\}$ be in community 1, nodes in $\{n/2+1,n/2+2,\ldots, n\}$ be in community 2, and $P$ be
\begin{flalign*}
P_{4}&=\begin{bmatrix}
    1&0.1\\
    0.1&0.8
\end{bmatrix} \mathrm{~or~}
P_{5}=\begin{bmatrix}
    1&-0.1\\
    -0.1&-0.8
\end{bmatrix}.
\end{flalign*}

\emph{Set-up 1}: When $A(i,j)\sim \mathrm{Bernoulli}(\Omega(i,j))$ for $i,j\in[n]$, set $n=20,\rho=0.8$ and $P$ as $P_{4}$. Panel (a) of Figure \ref{AF} shows an adjacency matrix generated by (DFM+$G(n,p)$) for this set-up, where we also report error rates of all methods for this adjacency matrix in Table \ref{TableSetUp}. With given $\hat{A}_{\mathcal{A}}$ and known $Z$, readers can apply these methods to $\hat{A}_{\mathcal{A}}$ to check their effectiveness.

\emph{Set-up 2}: When $A(i,j)\sim \mathrm{Normal}(\Omega(i,j),\sigma^{2}_{A})$ for $i,j\in[n]$, set $n=12,\sigma^{2}_{A}=1, \rho=2$ and $P$ as $P_{5}$. Panel (b) of Figure \ref{AF} shows an $\hat{A}_{\mathcal{A}}$ generated by (DFM+$G(n,p)$) for this set-up.

\emph{Set-up 3}: When $A(i,j)\sim \mathrm{Binomial}(m,\Omega(i,j)/m)$ for $i,j\in[n]$, set $n=20, \rho=2, m=4$, and $P$ as $P_{4}$. Panel (c) of Figure \ref{AF} shows an $\hat{A}_{\mathcal{A}}$ generated by (DFM+$G(n,p)$) for this set-up.

\emph{Set-up 4}: When $A(i,j)\sim \mathrm{Poisson}(\Omega(i,j))$ for $i,j\in[n]$, set $n=20, \rho=20$ and $P$ as $P_{4}$. Panel (d) of Figure \ref{AF} shows an $\hat{A}_{\mathcal{A}}$ generated by (DFM+$G(n,p)$) for this set-up.

\emph{Set-up 5}: When $A(i,j)\sim \mathrm{Exponential}(\frac{1}{\Omega(i,j)})$ for $i,j\in[n]$, set $n=12, \rho=20$ and $P$ as $P_{4}$. Panel (e) of Figure \ref{AF} shows an $\hat{A}_{\mathcal{A}}$ generated by (DFM+$G(n,p)$) for this set-up.

\emph{Set-up 6}: For signed networks when $\mathbb{P}(A(i,j)=1)=\frac{1+\Omega(i,j)}{2}$ and $\mathbb{P}(A(i,j)=-1)=\frac{1-\Omega(i,j)}{2}$ for $i,j\in[n]$, set $n=24, \rho=0.8$ and $P$ as $P_{5}$. Panel (f) of Figure \ref{AF} shows an $\hat{A}_{\mathcal{A}}$ generated by (DFM+$G(n,p)$) for this set-up.
\end{Remark}
\subsection{Performance on empirical networks}
\begin{table}[h!]
\footnotesize
	\centering
	\caption{Basic information and summarized statistics of weighted networks with known community information used in this paper.}
	\label{realdata}
	\resizebox{\columnwidth}{!}{
	\begin{tabular}{cccccccccccc}
\hline\hline&Source&Node meaning&Edge meaning&$n$&$K$&$\mathrm{max}_{i,j}A(i,j)$&$\mathrm{min}_{i,j}A(i,j)$&\#Edges&\%Positive edges\\
\hline
Karate club&\cite{karate}&Member&Tie&34&2&7&0&78&100\%\\
Gahuku-Gama subtribes&\cite{read1954cultures}&Tribe&Friendship&16&3&1&-1&58&50\%\\
\hline\hline
\end{tabular}}
\end{table}
\begin{figure}
\centering
\subfigure[]{\includegraphics[width=0.325\textwidth]{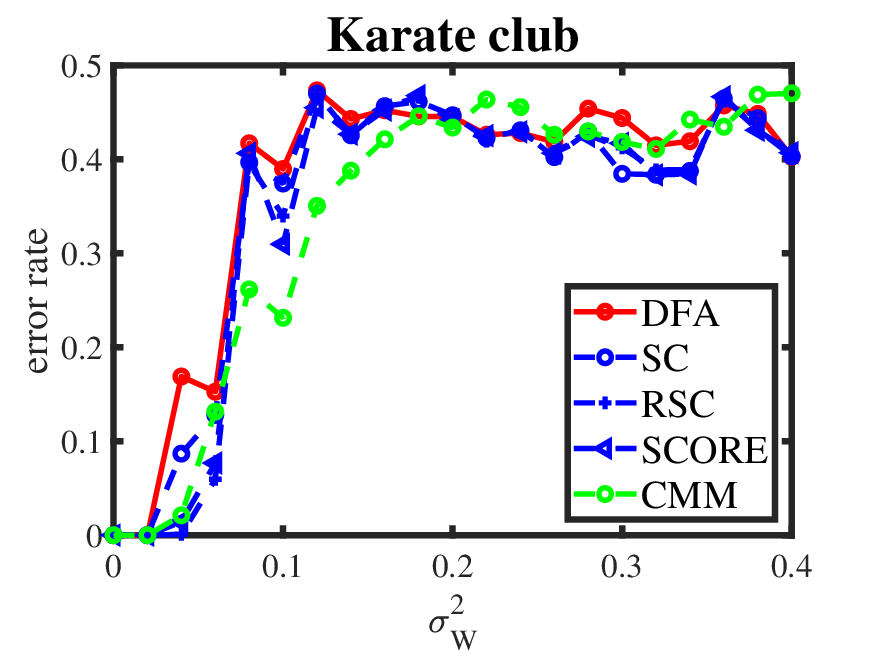}}
\subfigure[]{\includegraphics[width=0.325\textwidth]{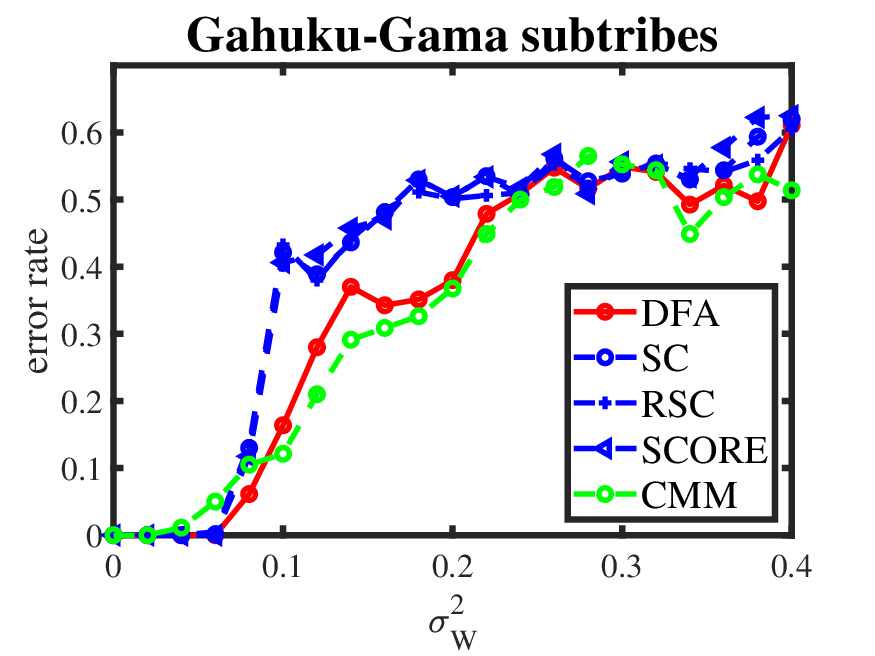}}
\caption{Error rates on the Karate club and Gahuku-Gama subtribes networks.}
\label{Real} 
\end{figure}

\begin{figure}
\centering
\subfigure[]{\includegraphics[width=0.4\textwidth]{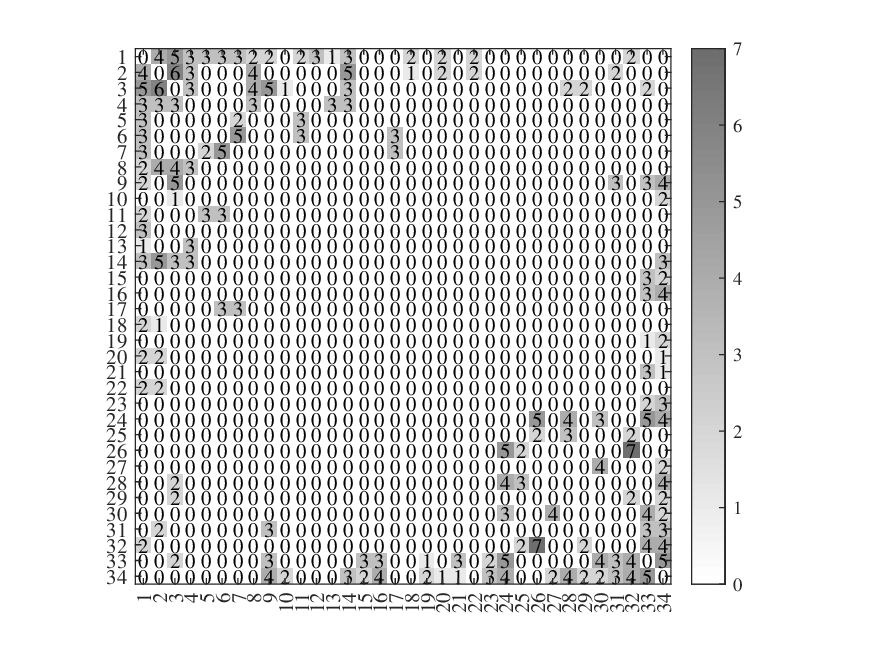}}
\subfigure[]{\includegraphics[width=0.4\textwidth]{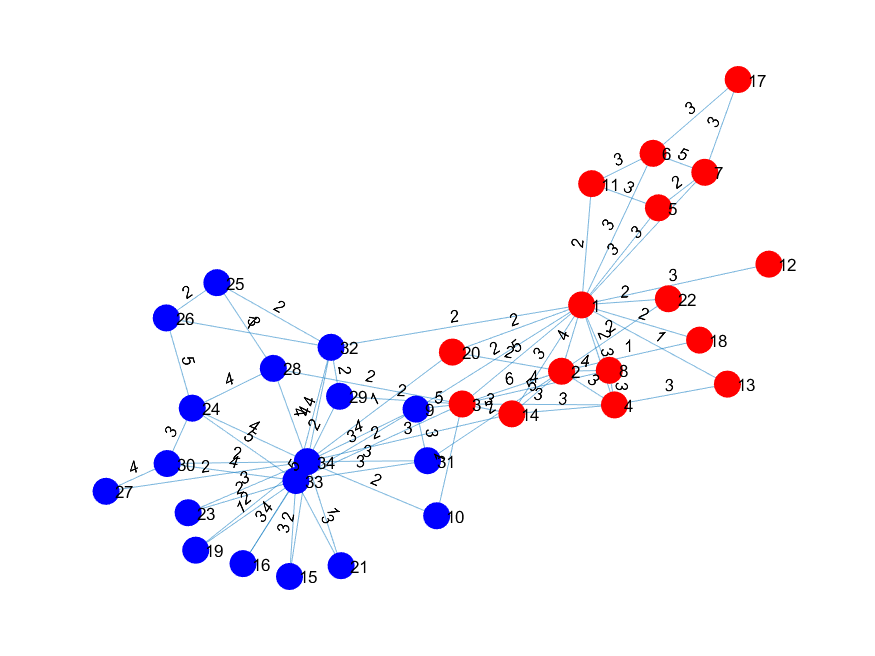}}
\subfigure[]{\includegraphics[width=0.4\textwidth]{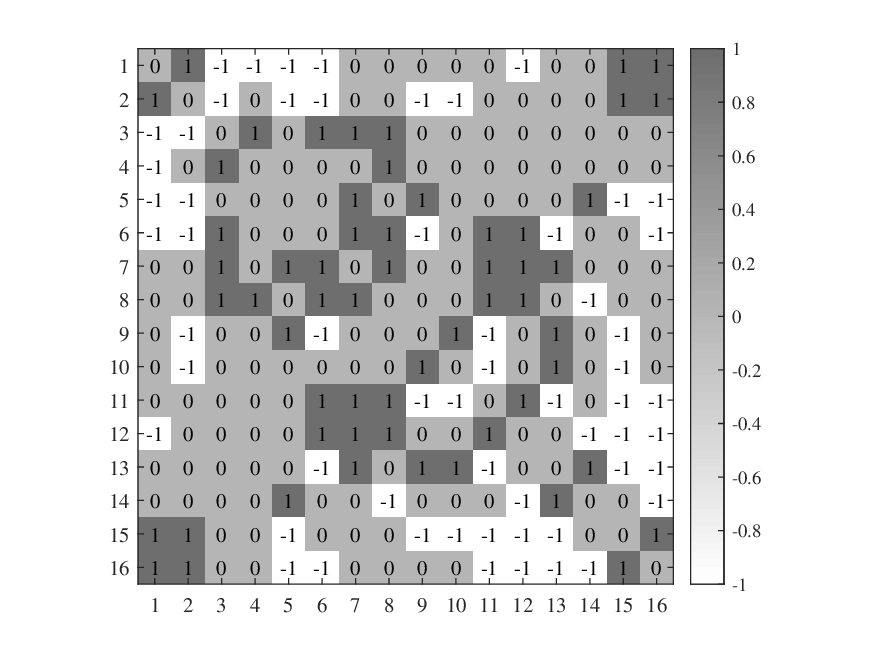}}
\subfigure[]{\includegraphics[width=0.4\textwidth]{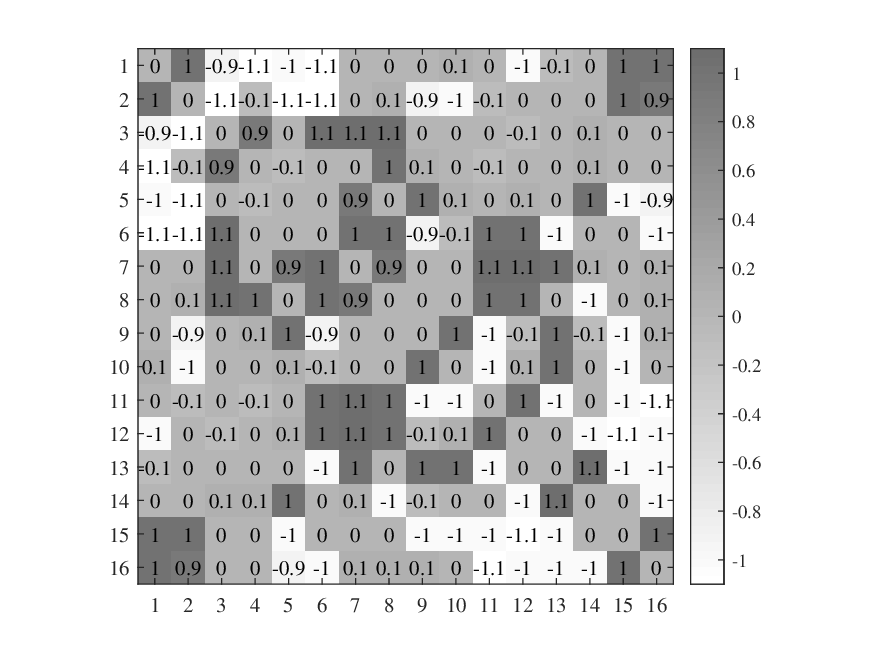}}
\subfigure[]{\includegraphics[width=0.4\textwidth]{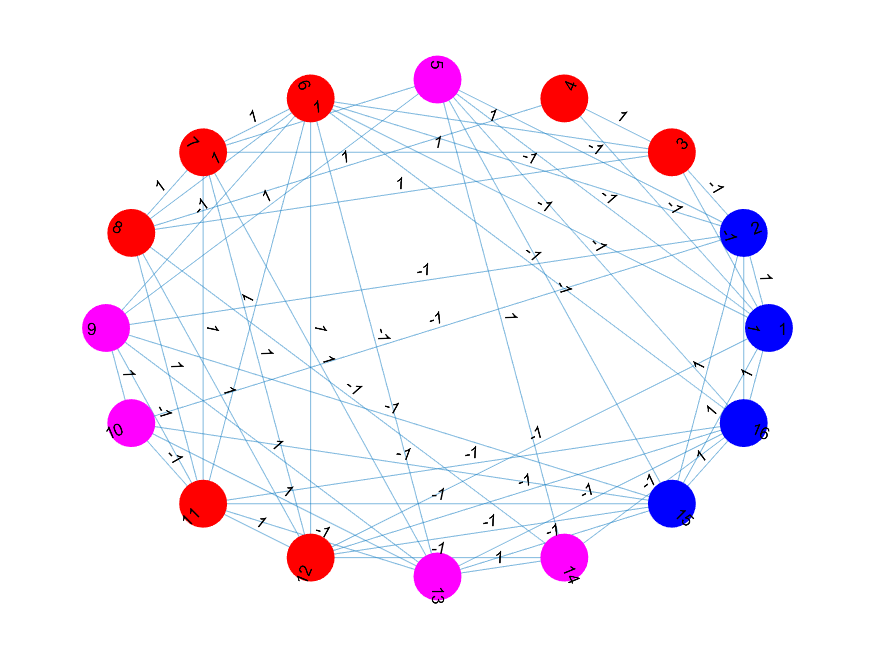}}
\subfigure[]{\includegraphics[width=0.4\textwidth]{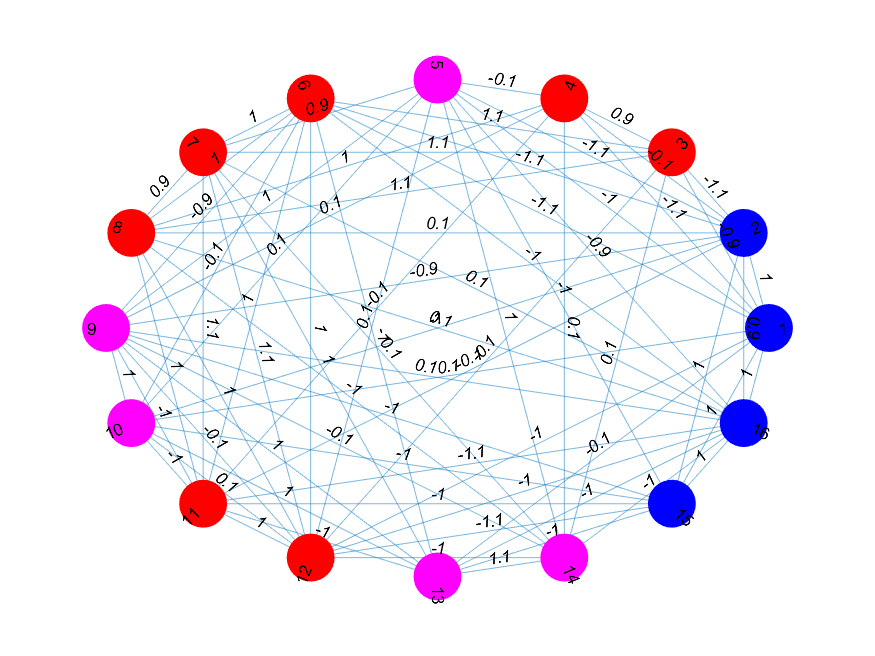}}
\caption{Panel (a): $A$ for the Karate club network. Panel (b): communities detected by DFA for the Karate club network without considering noise. Panel (c): $A$ for the Gahuku-Gama subtribes network. Panel (d): $\hat{A}$ for the Gahuku-Gama subtribes network when $W\sim\mathrm{Normal}(0,0.05)$. Panel (e): communities detected by DFA for the Gahuku-Gama subtribes network when using $A$. Panel (f): communities detected by DFA for the Gahuku-Gama subtribes network when using $\hat{A}$ in panel (d). For panels (b), (e), and (f), DFA's error rates are 0, and colors indicate communities. In panels (d) and (f), we keep $A$'s elements in one decimal for visualization.}
\label{RealNetworks} 
\end{figure}
This subsection studies the performances of these methods on two real-world weighted networks with known community information, the Zachary's Karate club weighted network (Karate club for short) and the Gahuku-Gama subtribes network. The Karate club network can be downloaded from \url{http://vlado.fmf.uni-lj.si/pub/networks/data/ucinet/ucidata.htm#kazalo}. The Gahuku-Gama subtribes network can be downloaded from \url{http://konect.cc/networks/ucidata-gama/}(see also \cite{kunegis2013konect}). True community information of the Karate club is known and can be found from \url{http://www-personal.umich.edu/~mejn/netdata/}. For Gahuku-Gama subtribes, we use node labels provided in Figure 9 (b) from \cite{yang2007community} as ground truth. Table \ref{realdata} summarizes basic information for these two weighted networks.

To study the behaviors of all methods when there exists noise, we compute the observed matrix $\hat{A}=A+W$ where $W\sim\mathrm{Normal}(0,\sigma^{2}_{W})$ and let $\sigma^{2}_{W}$ range in $\{0,0.02,0.04,\ldots,0.4\}$. Note that, when $\sigma^{2}_{W}=0$, the observed adjacency matrix $\hat{A}$ is $A$. For each $\sigma^{2}_{W}$, we report the averaged error rate among 50 repetitions for every method. Figure \ref{Real} records error rates of these methods on Karate club and Gahuku-Gama subtribes networks. We see that all methods perform better when $\sigma^{2}_{W}$ decreases, and these methods have stable performances when $\hat{A}\neq A$ under the case that $\sigma^{2}_{W}$ is small. Meanwhile, when $W$ is a zero matrix, the error rates of all methods for these two weighted networks are 0. Therefore, these methods enjoy performance stability when $A$ is slightly perturbed by some noise, and this is consistent with our theoretical and simulation results. For visibility, Figure \ref{RealNetworks} depicts $A$, $\hat{A}$, and communities returned by DFA for Karate club and Gahuku-Gama subtribes networks.
\section{Conclusions}
In this paper, we introduced the Distribution-Free Model for weighted networks. DFM provides powerful and exploratory tools for network analysis and it can model networks generated under various distributions. To model real-world weighted networks with missing edges, we summarize a four-step data generation process. Our experiments show that some benchmark community detection methods can satisfactorily detect communities for adjacency matrices generated by our four-step data generation process under different distributions.

DFM is a generative model like SBM, and it can be studied and extended in many potential ways. For example, estimating the number of communities $K$ under DFM for different distribution $\mathcal{F}$ is an interesting and challenging problem; building theoretical guarantees of spectral algorithms via applications of the Laplacian matrix studied in \cite{rohe2011spectral, RSC,joseph2016impact,su2019strong} to fit our DFM is an interesting topic; similar to the exact recovery problem under SBM studied in \cite{abbe2016exact,hajek2016achieving,abbe2018proof}, the exact recovery problem under DFM is interesting; estimating communities with theoretical guarantees under DFM by semidefinite optimization and convex optimization approaches \cite{guedon2016community,amini2018semidefinite,chen2018convexified,fei2018exponential,li2021convex} is appealing; extending DFM to hierarchical weighted networks \cite{lyzinski2016community} is also interesting. We leave studies of these problems for our future work.
\section*{Acknowledgements}
This research was funded by the High level personal project of Jiangsu Province [JSSCBS20211218].

\bibliographystyle{ptephy}
\bibliography{refDFM}
\appendix
\section{Proofs under DFM}
\subsection{Proof of Proposition \ref{idDFM}}
\begin{proof}
By Lemma \ref{GenUrUc}, since $U=ZB$, we have $ZB=\tilde{Z}B$. Let $\tilde{\ell}$ be the $n\times 1$ label vector obtained from $\tilde{Z}$, we have $e'_{i}ZB=Z(i,:)B=B(\ell(i),:)=B(\tilde{\ell}(i),:)$, which gives $\ell(i)=\tilde{\ell}(i)$ for $i\in[n]$. Hence, $Z=\tilde{Z}$. Since $\Omega(\mathcal{I},\mathcal{I})=\rho Z(\mathcal{I},:)PZ'(\mathcal{I},:)=\rho P=\rho\tilde{Z}(\mathcal{I},:)\tilde{P}\tilde{Z}'(\mathcal{I},:)=\rho\tilde{P}$, we have $P=\tilde{P}$, and this proposition follows.
\end{proof}
\subsection{Proof of Lemma \ref{GenUrUc}}
\begin{proof}
Since $\Omega=\rho ZPZ'=U\Lambda U'$ and $U'U=I_{K_{0}}$, we have $U=\rho ZPZ'U\Lambda^{-1}=Z\rho PZ'U\Lambda^{-1}$, i.e., $B=\rho PZ'U\Lambda^{-1}$. Since $U=ZB$, we have $U(\mathcal{I},:)=Z(\mathcal{I},:)B=B$, which gives $B=U(\mathcal{I},:)$.
\end{proof}
\subsection{Proof of Lemma \ref{BoundAOmega}}
\begin{proof}
We apply the Theorem 1.4 (the Matrix Bernstein) of \cite{tropp2012user} to bound $\|A-\Omega\|$. Write this theorem below
\begin{Theorem}\label{Bern}
Consider a finite sequence $\{X_{k}\}$ of independent, random, self-adjoint matrices with dimension $d$. Assume that each random matrix satisfies
\begin{align*}
\mathbb{E}[X_{k}]=0, \mathrm{and~}\|X_{k}\|\leq R~\mathrm{almost~surely}.
\end{align*}
Then, for all $t\geq 0$,
\begin{align*}
\mathbb{P}(\|\sum_{k}X_{k}\|\geq t)\leq d\cdot \mathrm{exp}(\frac{-t^{2}/2}{\sigma^{2}+Rt/3}),
\end{align*}
where $\sigma^{2}:=\|\sum_{k}\mathbb{E}(X^{2}_{k})\|$.
\end{Theorem}
It should be emphasized that Theorem \ref{Bern} also has no constraint on the distribution of $\{X_{k}\}$ as long as $\{X_{k}\}$ are finite independent, random, self-adjoint matrices with expectation 0 and $\|X_{k}\|\leq R$, and this is the reason we can apply this theorem to bound $\|\hat{A}-\Omega\|$ without violating DFM's distribution-free property.

Let $e_{i}$ be an $n\times 1$ vector with $e_{i}(i)=1$ and $0$ elsewhere for $i\in[n]$. Set $H=\hat{A}-\Omega$, then $H=\sum_{i=1}^{n}\sum_{j=1}^{n}H(i,j)e_{i}e'_{j}$. Set $H^{(i,j)}=H(i,j)e_{i}e'_{j}$. Since $\mathbb{E}[H(i,j)]=\mathbb{E}[\hat{A}(i,j)-\Omega(i,j)]=\mathbb{E}[A(i,j)+W(i,j)-\Omega(i,j)]=\mathbb{E}[A(i,j)]-\Omega(i,j)+\mathbb{E}[W(i,j)]=0$, we have $\mathbb{E}[H^{(i,j)}]=0$ and
\begin{align*}
\|H^{(i,j)}\|&=\|(\hat{A}(i,j)-\Omega(i,j))e_{i}e'_{j}\|=|\hat{A}(i,j)-\Omega(i,j)|\|e_{i}e'_{j}\|=|\hat{A}(i,j)-\Omega(i,j)|\\
&=|A(i,j)+W(i,j)-\Omega(i,j)|\leq\tau,
\end{align*}
i.e., $R=\tau$.

For the variance term $\sigma^{2}=\|\sum_{i=1}^{n}\sum_{j=1}^{n}\mathbb{E}[H^{(i,j)}(H^{(i,j)})']\|$, since $\mathbb{E}[H^{2}(i,j)]=\mathbb{E}[(\hat{A}(i,j)-\Omega(i,j))^{2}]=\mathbb{E}[(A(i,j)+W(i,j)-\Omega(i,j))^{2}]=\mathbb{E}[(A(i,j)-\Omega(i,j))^{2}]+2\mathbb{E}[W(i,j)(A(i,j)-\Omega(i,j))]+\mathbb{E}[W^{2}(i,j)]=\mathrm{Var}(A(i,j))+\mathbb{E}[(W(i,j)-\mathbb{E}[W(i,j)])^{2}]=\mathrm{Var}(A(i,j))+\mathrm{Var}(W(i,j))\leq \gamma\rho+\sigma^{2}_{W}$ by Equation (\ref{DefinW}), we have
\begin{align*}
\|\sum_{i=1}^{n}\sum_{j=1}^{n}\mathbb{E}(H^{(i,j)}(H^{(i,j)})')\|=\|\sum_{i=1}^{n}\sum_{j=1}^{n}\mathbb{E}(H^{2}(i,j))e_{i}e'_{j}e_{j}e'_{i}\|=\|\sum_{i=1}^{n}\sum_{j=1}^{n}\mathbb{E}(H^{2}(i,j))e_{i}e'_{i}\|\leq \gamma\rho n+\sigma^{2}_{W}n.
\end{align*}
So, we have $\sigma^{2}\leq \gamma\rho n+\sigma^{2}_{W}n$. Set $t=\frac{\alpha+1+\sqrt{\alpha^{2}+20\alpha+19}}{3}\sqrt{(\gamma\rho n+\sigma^{2}_{W}n)\mathrm{log}(n)}$. By Theorem \ref{Bern}, we have
\begin{align*}
&\mathbb{P}(\|H\|\geq t)\leq n\mathrm{exp}(-\frac{t^{2}/2}{\sigma^{2}+\frac{Rt}{3}})\leq n\mathrm{exp}(-\frac{t^{2}/2}{(\gamma\rho n+\sigma^{2}_{W}n)+\frac{Rt}{3}})\\
&=n\mathrm{exp}(-(\alpha+1)\mathrm{log}(n)\cdot \frac{1}{\frac{18}{(\sqrt{\alpha+19}+\sqrt{\alpha+1})^{2}}+\frac{2\sqrt{\alpha+1}}{\sqrt{\alpha+19}+\sqrt{\alpha+1}}\sqrt{\frac{R^{2}\mathrm{log}(n)}{\gamma\rho n+\sigma^{2}_{W}n}}})\\
&\leq n\mathrm{exp}(-(\alpha+1)\mathrm{log}(n))=\frac{1}{n^{\alpha}},
\end{align*}
where the last inequality comes from Assumption \ref{assumesparsity} such that $\frac{18}{(\sqrt{\alpha+19}+\sqrt{\alpha+1})^{2}}+\frac{2\sqrt{\alpha+1}}{\sqrt{\alpha+19}+\sqrt{\alpha+1}}\sqrt{\frac{R^{2}\mathrm{log}(n)}{\gamma\rho n+\sigma^{2}_{W}n}}\leq\frac{18}{(\sqrt{\alpha+19}+\sqrt{\alpha+1})^{2}}+\frac{2\sqrt{\alpha+1}}{\sqrt{\alpha+19}+\sqrt{\alpha+1}}=1$. Thus, the claim follows.
\end{proof}
\subsection{Proof of Theorem \ref{MainDFA}}
\begin{proof}
First, we provide a general lower bound of $\sigma_{K_{0}}(\Omega)$ under DFM, and such a lower bound is directly related to the sparsity parameter $\rho$ and the separation parameter $\sigma_{K_{0}}(P)$.
\begin{Lemma}\label{svdK}
Under $DFM_{n}(K,P,Z,\rho,\mathcal{F})$, we have
\begin{align*}
\sigma_{K_{0}}(\Omega)\geq\rho\sigma_{K_{0}}(P)n_{K_{0}}. \end{align*}
\end{Lemma}
\begin{proof}
For $\sigma_{K}(\Omega)$, we have
\begin{align*}
\sigma^{2}_{K_{0}}(\Omega)&=\rho^{2}\lambda_{K_{0}}(\Omega\Omega')=\rho^{2}\lambda_{K_{0}}(ZPZ'ZPZ')=\rho^{2}\lambda_{K_{0}}(Z'ZPZ'ZP')\geq \rho^{2}\lambda_{K_{0}}(Z'Z)\lambda_{K_{0}}(PZ'ZP)\\
&=\rho^{2}\lambda_{K_{0}}(Z'Z)\lambda_{K_{0}}(Z'ZP^{2})\geq\rho^{2}\lambda^{2}_{K_{0}}(Z'Z)\lambda_{K_{0}}(P^{2})\geq\rho^{2}\sigma^{2}_{K_{0}}(P)n^{2}_{K_{0}},
\end{align*}
where we have used the fact for any matrices $X, Y$, the nonzero eigenvalues of $XY$ are the same as the nonzero eigenvalues of $YX$.
\end{proof}
Since bound in Lemma 5.1 of \cite{lei2015consistency} is irrelevant with distribution, we can apply Lemma 5.1 of \cite{lei2015consistency} to bound $\hat{U}$ and $U$  without violating DFM's distribution-free property. By Lemma 5.1 of \cite{lei2015consistency}, there exists a $K_{0}\times K_{0}$ orthogonal matrix $Q$ such that
\begin{align*}
\|\hat{U}Q-U\|_{F}\leq \frac{2\sqrt{2K_{0}}\|\hat{A}-\Omega\|}{\sigma_{K_{0}}(\Omega)}.
\end{align*}
Under $DFM(n,K_{0},K,P,Z,\rho)$, by Lemma \ref{svdK}, we have $\sigma_{K_{0}}(\Omega)\geq\rho\sigma_{K_{0}}(P)\sigma^{2}_{K_{0}}(Z)=\rho\sigma_{K_{0}}(P)n_{K_{0}}$,
which gives
\begin{align*}
\|\hat{U}Q-U\|_{F}\leq \frac{2\sqrt{2K_{0}}\|\hat{A}-\Omega\|}{\sigma_{K_{0}}(P)\rho n_{K_{0}}}.
\end{align*}
Let $\varsigma>0$ be a small quantity, by Lemma 2 in \cite{joseph2016impact} where this lemma is also distribution-free if one has
\begin{align}\label{holdGISBM}
\frac{\sqrt{K}}{\varsigma}\|U-\hat{U}Q\|_{F}(\frac{1}{\sqrt{n_{k}}}+\frac{1}{\sqrt{n_{l}}})\leq \|B(k,:)-B(l,:)\|_{F}, \mathrm{~for~each~}1\leq k\neq l\leq K,
\end{align}
then the clustering error $\hat{f}=O(\varsigma^{2})$. Recall that we set $\delta=\mathrm{min}_{k\neq l}\|B(k,:)-B(l,:)\|_{F}$ to measure the minimum center separation of $B$. Setting $\varsigma=\frac{2}{\delta}\sqrt{\frac{K}{n_{\mathrm{min}}}}\|U-\hat{U}Q\|_{F}$ makes Equation (\ref{holdGISBM}) hold for all $1\leq k\neq l\leq K$. Then we have $\hat{f}=O(\varsigma^{2})=O(\frac{K\|U-\hat{U}Q\|^{2}_{F}}{\delta^{2}n_{\mathrm{min}}})$. By the upper bound of $\|\hat{U}Q-U\|_{F}$, we have
\begin{align*}
\hat{f}=O(\frac{K_{0}K\|
\hat{A}-\Omega\|^{2}}{\sigma^{2}_{K_{0}}(P)\rho^{2}\delta^{2}n^{2}_{K_{0}}n_{\mathrm{min}}}).
\end{align*}
Especially, for the special case when $K_{0}=K$, $\delta\geq \sqrt{\frac{2}{n_{\mathrm{max}}}}$ under DFM by Lemma 2.1 (a distribution-free lemma) of \cite{lei2015consistency} . Note that, it is a challenge to obtain a positive lower bound of $\delta$ if $K_{0}\neq K$. Meanwhile, when $K_{0}=K$, $n_{K_{0}}=n_{\mathrm{min}}$, which gives
\begin{align*}
\hat{f}=O(\frac{K^{2}\|\hat{A}-\Omega\|^{2}n_{\mathrm{max}}}{\sigma^{2}_{K}(P)\rho^{2}n^{3}_{\mathrm{min}}}).
\end{align*}
When $K_{0}=K=O(1)$ and $\frac{n_{\mathrm{max}}}{n_{\mathrm{min}}}=O(1)$, the last statement of this theorem follows immediately by basic algebra. Finally, use the theoretical bound of Lemma \ref{BoundAOmega} to replace $\|\hat{A}-\Omega\|$, and this theorem follows.
\end{proof}
\end{document}